\documentclass[runningheads]{llncs}
\AtBeginDocument{%
  }

\usepackage[utf8]{inputenc}
\usepackage[T1]{fontenc}
\usepackage{textcomp}
\usepackage{graphicx}
\usepackage{subcaption} %for sub-figures 
\usepackage{float}%
\usepackage[table]{xcolor}
\usepackage{xcolor}
\usepackage{color}
\usepackage{xspace,tcolorbox}
\usepackage{tikz}
\usetikzlibrary{shapes,arrows,patterns,positioning} 
\usepackage{caption}
\usepackage[normalem]{ulem}
\usepackage{hyperref} 
\usepackage{tabularx}

\usepackage{amsmath}
\usepackage{amssymb}
\usepackage{empheq}
\usepackage[mathscr]{euscript}
\DeclareSymbolFont{rsfs}{U}{rsfs}{m}{n}
\DeclareSymbolFontAlphabet{\mathscrsfs}{rsfs}
\usepackage{mathtools}
\usepackage{stmaryrd} % for \bigparallel

\usepackage{dutchcal} %\mathcal: also small letters. \mathbcal = bold ones. 
\usepackage{pgfplots}
\pgfplotsset{compat=newest}

\usepackage[utf8]{inputenc}
\usepackage[T1]{fontenc}
\usepackage{textcomp,mathcomp}
\usepackage{xcolor,colortbl} %colortbl is for Table background color

\usepackage{chngcntr} %To change figure/table numbers in appendix. It includes the command \counterwithin.

\newcommand{\halfcheckmark}{{$\checkmark$}\textsuperscript{\textcolor{black}{\kern-0.53em{\bf--}}}}

\usepackage{threeparttable}

\usepackage{booktabs}
\usepackage{nicematrix}

\usepackage{multirow}
\usepackage[ruled,linesnumbered]{algorithm2e}

\usepackage{breakcites}
\usepackage{enumitem}

\usepackage{placeins} %for \FloatBarrier

\usepackage{wrapfig} %To wrap figure
\usepackage{subcaption} %for sub-figures 

\newcommand{\optional}[1]{{\color{violet}#1}}

\newcommand{\eyasu}[1]{{\color{red!}{E: #1}}}
\newcommand{\howard}[1]{{\color{red!}{H: #1}}}

\definecolor{rulebasedbg}{RGB}{222, 235, 247}
\definecolor{scorebasedbg}{RGB}{226, 240, 217}

\usepackage{xargs} %Use more than one optional parameter in a new commands
\usepackage[colorinlistoftodos,prependcaption,textsize=tiny]{todonotes}
\newcommandx{\commentt}[2][1=]{\todo[linecolor=red,backgroundcolor=red!25,bordercolor=red,#1]{#2}}

\makeatletter
\newlength\min@xx
\newcommand*\xxrightarrow[1]{\begingroup
  \settowidth\min@xx{$\m@th\scriptstyle#1$}
  \@xxrightarrow}
\newcommand*\@xxrightarrow[2][]{
  \sbox8{$\m@th\scriptstyle#1$}  % subscript
  \ifdim\wd8>\min@xx \min@xx=\wd8 \fi
  \sbox8{$\m@th\scriptstyle#2$} % superscript
  \ifdim\wd8>\min@xx \min@xx=\wd8 \fi
  \xrightarrow[{\mathmakebox[\min@xx]{\scriptstyle#1}}]
    {\mathmakebox[\min@xx]{\scriptstyle#2}}
  \endgroup}
\makeatother

\makeatletter
\newcommand*\xxleftarrow[1]{\begingroup
  \settowidth\min@xx{$\m@th\scriptstyle#1$}
  \@xxleftarrow}
\newcommand*\@xxleftarrow[2][]{
  \sbox8{$\m@th\scriptstyle#1$}  % subscript
  \ifdim\wd8>\min@xx \min@xx=\wd8 \fi
  \sbox8{$\m@th\scriptstyle#2$} % superscript
  \ifdim\wd8>\min@xx \min@xx=\wd8 \fi
  \xleftarrow[{\mathmakebox[\min@xx]{\scriptstyle#1}}]
    {\mathmakebox[\min@xx]{\scriptstyle#2}}
  \endgroup}
\makeatother

\usepackage[left=3.6cm, right = 3.6cm, top=3.6cm, bottom=3.6cm, asymmetric]{geometry}

\begin{document}
%\title{Reconstructing Biometric Features from Biometric Traces}
%\title{LACBBA/LARBBA/LaCTuR/LaCR{B/BB}A: LLM-Assisted Contextual {Trust} Reasoning for Behavioral Biometric Authentication}
\title{LABBA: LLM-Assisted Behavioral Biometric Authentication}

%\author{No Author Given}
%\institute{No Institute Given}

 \author{Howard Halim \and Eyasu Getahun Chekole \and Jianying Zhou}
\institute{Singapore University of Technology and Design, Singapore
\email{\{howard\_halim,eyasu\_chekole,jianying\_zhou\}@sutd.edu.sg}}

\maketitle
\begin{abstract}
Behavioral biometric authentication has emerged as a promising complement to traditional physiological biometrics by leveraging dynamic human characteristics such as typing patterns, gait, speech articulation, touchscreen interactions, and airflow-based behaviors. These modalities offer advantages including continuous authentication, inherent liveness detection, and greater resistance to credential theft and various types of spoofing attacks. However, existing systems remain limited in adaptability, contextual awareness, and semantic reasoning as they primarily rely on similarity metrics, static thresholds, or supervised classifiers that perform well only in controlled environments. As a result, they often struggle with real-world conditions such as behavioral variability, environmental changes, contextual changes, and adversarial manipulations.

%To address these limitations, we systematically integrate Large Language Model (LLM)-based contextual reasoning into behavioral biometric authentication. Unlike conventional machine learning approaches, which are primarily optimized for pattern recognition, LLMs offer advanced capabilities in contextual inference, sequential reasoning, multimodal understanding, and adaptive decision-making. Leveraging these capabilities, we systematically integrate a lightweight LLM-based contextual reasoning with a novel phone-blowing behavioral biometric technique to establish a context-aware and cognitively driven authentication framework. To this end, we first design the biometric technique and collect blow-acoustic signals from participants from which spectral and temporal features are systematically extracted. The extracted features are jointly encoded into a symbolic representation, which is then transformed into a structured prompt using prompt engineering, which is then processed by the LLM contextual reasoning module to make prompt interpretation and generate context-aware authentication decisions. Therefore, the proposed framework dynamically interprets blow-acoustic behavioral biometric signals across diverse behavioral and contextual conditions, thereby enabling adaptive and context-aware authentication decisions. By replacing rigid static-threshold mechanisms with LLM-driven contextual reasoning, the framework enhances authentication robustness under dynamic operating environments and behavioral variability.

To address these limitations, we systematically integrate Large Language Model (LLM)-based contextual reasoning into behavioral biometric authentication. Unlike conventional machine learning methods, %focused primarily on pattern recognition, 
LLMs provide capabilities in contextual inference, sequential reasoning, multimodal understanding, and adaptive decision-making. Building on these strengths, we %systematically 
integrate a lightweight LLM-based reasoning module with a novel phone-blowing behavioral biometric technique to develop a context-aware authentication framework. To this end, we first collect blow-acoustic signals from participants and systematically extract spectral and temporal features. %These features are aggregated and encoded into symbolic representations and transformed into structured prompts through prompt engineering. 
%\addnew{
These features are aggregated and converted into comparable similarity representations, using either discrete rule-based categories or continuous normalized scores, which are transformed into structured prompts through prompt engineering. %} 
These prompts are processed by the LLM reasoning module to generate context-aware authentication decisions. 

%To evaluate the effectiveness of the proposed framework, blow-acoustic signals are collected from 50 participants under diverse behavioral and contextual conditions. Temporal and spectral blow-acoustic features are extracted and integrated with the LLM reasoning engine to support adaptive authentication decision-making. Experimental results demonstrate that the proposed framework achieves an authentication accuracy of 99.43\%, significantly outperforming conventional static threshold-based approaches while exhibiting strong robustness against dynamic real-world conditions with behavioral and contextual variability.

%We experimentally evaluate the the proposed framework by collecting blow-acoustic data from 50 participants under diverse behavioral and contextual conditions. %Temporal and spectral features were extracted and integrated with an LLM-based reasoning engine to enable adaptive authentication decisions. 
%Experimental results demonstrate that 
%\howard{which result should we present here, the old one has a slightly higher accuracy but lower EER.}
We experimentally evaluate the proposed framework using blow-acoustic data collected from 50 participants under diverse behavioral and contextual conditions. The proposed LLM-assisted framework achieves 99.46\% and 99.38\% authentication accuracy in the rule-based and score-based %evaluation 
modes, respectively, outperforming its non-LLM counterpart while maintaining strong robustness under dynamic real-world conditions. By replacing rigid static threshold-based mechanisms with LLM-driven contextual reasoning, the framework enhances behavioral biometric authentication robustness against dynamic environmental conditions and behavioral variability.
\end{abstract}

%\begin{key}

\section{Introduction}
\label{introduction}

The rapid growth of digital services, mobile computing, and interconnected %cyber-physical systems 
critical infrastructures has significantly increased the demand for secure, seamless, and user-friendly authentication mechanisms. Conventional authentication methods such as passwords, PINs, and security tokens suffer from several well-known limitations, including susceptibility to credential theft, phishing attacks, replay attacks, shoulder surfing, and poor usability due to credential memorization requirements \cite{chimuco2023secure,angcsur2025}. As a result, biometric authentication has emerged as a promising alternative for identity verification by leveraging intrinsic human characteristics that are difficult to replicate or share \cite{jain2005biometric,sharma2022reliable,halim2025blowprint}.

Traditional biometric systems primarily rely on physiological characteristics \cite{alsaadi2015physiological} such as fingerprints, facial appearance, iris patterns, and palmprints. While these modalities provide strong discriminative capability, they often suffer from several practical limitations, including vulnerability to spoofing attacks, lack of continuous authentication capability, sensor dependency, and privacy concerns associated with the permanent nature of physiological traits. In contrast, behavioral biometrics \cite{finnegan2024utility,fereidooni2023authentisense,khan2024mouse} has gained increasing attention as a complementary and, in some cases, alternative authentication paradigm that utilizes dynamic human behaviors such as typing dynamics, gait patterns, speech articulation, touchscreen interactions, mouse movement patterns, and airflow or breathing behaviors.

%% Optional paragraph
%%\optional{Behavioral biometrics offer several important advantages over conventional physiological biometrics. Since behavioral characteristics are generated dynamically during user interaction, they inherently support continuous and passive authentication without requiring explicit user intervention. Moreover, behavioral signals often exhibit implicit liveness characteristics, making them more resistant to simple spoofing attempts. Their unobtrusive nature also improves usability and user experience in mobile and pervasive computing environments. Consequently, behavioral biometric authentication has become increasingly relevant for applications including smartphone security, wearable devices, online banking, remote access systems, healthcare platforms, and intelligent human-computer interaction systems.}

Despite substantial advances in behavioral biometrics, existing approaches remain fundamentally constrained by their reliance on static template or signal matching paradigms. %%Cite
Most such systems employ similarity-based verification models \cite{shadman2025keystroke}, fixed threshold mechanisms \cite{avola2024signal}, statistical classifiers \cite{kokal2022exploration}, or supervised machine learning \cite{madavarapu2024behavioral} approaches that operate primarily at %the 
signal or feature levels. Although these techniques demonstrate promising performance under controlled experimental settings, their effectiveness often degrades significantly in real-world deployments characterized by natural behavioral variability \cite{subash2026adaptability,pisani2024study}, dynamic contextual conditions \cite{ayeswarya2025enhancing,solano2021combining}, and sophisticated adversarial threats \cite{solano2020scrap,qin2024improving}. 

One of the primary challenges in behavioral biometric authentication is the highly dynamic nature of human behavior \cite{subash2026adaptability,pisani2024study}. Unlike physiological biometrics, behavioral patterns naturally evolve due to emotional state, cognitive workload, fatigue, stress, environmental conditions, device posture, interaction context, and long-term behavioral drift. For example, a user’s typing rhythm may vary depending on stress level or hand posture, while speech and breathing behaviors may change due to environmental noise, health condition, or physical activity. Similarly, airflow-based biometric interactions collected using smartphone microphones may exhibit variability caused by device orientation, ambient acoustic interference, user fatigue, or contextual interaction patterns. Such behavioral variability frequently causes conventional static threshold-based authentication systems to suffer from increased false rejection and false acceptance rates.

In addition to natural behavioral variability, most existing behavioral biometric systems %their inability to 
do not incorporate contextual awareness into authentication decision-making \cite{ayeswarya2025enhancing,solano2021combining}. In practical deployments, authentication trustworthiness depends not only on biometric similarity scores but also on contextual factors such as device state, environmental conditions, user interaction history, temporal patterns, sensor reliability, and session-level behavioral consistency. However, most current biometric frameworks treat authentication as an isolated classification problem and fail to exploit contextual information that may significantly improve robustness and reliability. This limitation becomes particularly critical in continuous authentication and multimodal biometric systems where authentication conditions continuously evolve over time.
 
Another important limitation of behavioral biometric systems is their increasingly exposed to sophisticated adversarial threats \cite{solano2020scrap,qin2024improving}. Attackers may attempt to imitate legitimate user behavior, inject synthetic behavioral traces, replay recorded interaction signals, or manipulate contextual conditions to bypass authentication systems. Existing anti-spoofing approaches primarily rely on low-level signal analysis \cite{li2023hiding,souza2025improving} or statistical anomaly detection techniques \cite{gheorghian2024study,habib2023revisiting}, which often lack the ability to reason about long-term behavioral consistency, contextual coherence, or user intent. Consequently, conventional behavioral biometric systems frequently operate as static decision engines that make authentication decisions solely based on similarity scores without understanding the semantic and contextual meaning of observed user behavior.

Recent advances in Large Language Models (LLMs) \cite{chang2024llm,wang2026adareasoner} provide a promising opportunity to address %these limitations and redefine the future of behavioral biometric authentication. 
the static threshold-based decision making limitation of biometric systems. Because LLMs have demonstrated strong %reasoning and adaptability 
contextual inference, sequential reasoning, multimodal interpretation, semantic understanding, and adaptive decision-making capabilities across a wide range of domains. This has led to increasing interest in their integration with biometric systems \cite{mehta2026detecting,almehmadi2025cyber,piland2026generalist,deandres2024good,farmanifard2024chatgpt,rastogi2025vascular}. 
%Moreover, research on the use of LLMs in biometric authentication is emerging, 
However, the application of LLMs in existing biometric systems is mainly limited to physiological biometrics, particularly in image-based modalities such as face, iris, and vein recognition \cite{deandres2024good,farmanifard2024chatgpt,rastogi2025vascular}. %These studies demonstrate promising potential in leveraging LLMs for biometric comparison and reasoning tasks. 
Their application to behavioral biometrics, such as acoustic-based biometrics, remains largely unexplored.

In this work, we design an LLM-assisted behavioral biometric authentication framework that leverages the powerful reasoning capabilities of LLMs to enable context-aware and cognitively driven behavioral biometric authentication. %\howard{the following claim of novel phone-blowing behavioral biometric technique, this is similar to blowprint/blowlive, can we claim this novel?}
Specifically, we first design a novel phone-blowing-based behavioral biometric technique that authenticates users based on their blow-acoustic signals. This modality captures rich behavioral and contextual characteristics, making it well suited for LLM-based contextual reasoning. During authentication, users interact with smartphone microphones through personalized airflow patterns and blowing behaviors, generating signals influenced by a wide range of behavioral and environmental factors, such as blowing style, airflow dynamics, user posture, %fatigue, 
ambient noise, microphone placement, and device orientation. Conventional static threshold-based methods often struggle to maintain reliability under such dynamic conditions, motivating the integration of LLM-based reasoning.

%To achieve this, we first design the biometric authentication technique and collect blow-acoustic signals from participants, from which 
%Then, we first systematically extract spectral and temporal features from the collected blow acoustic signals. The temporal features are particularly under gone through different bandpass filtering and Gaussian smoothing processes to suppress noises and unwanted frequency components. The extracted and processed spectral and temporal features are then aggregated and encoded into a symbolic representation and subsequently transformed into a structured prompt through prompt engineering. The structured prompt is then processed by the LLM-based contextual reasoning module to perform prompt interpretation and generate context-aware authentication decisions.

To address this challenge, we systematically extract spectral and temporal features from the collected blow-acoustic signals. The temporal features are further processed using bandpass filtering and Gaussian smoothing to suppress noise and unwanted frequency components. %The extracted features are then aggregated and encoded into symbolic representations, and transformed into structured prompts through prompt engineering.
The extracted features are then aggregated and converted into comparable similarity representations, using either discrete rule-based categories or continuous normalized scores, and transformed into structured prompts through prompt engineering. These prompts are subsequently processed by the LLM-based contextual reasoning module to interpret behavioral patterns and generate context-aware authentication decisions.

%Consequently, the proposed framework dynamically interprets blow-acoustic behavioral biometric signals across diverse behavioral and contextual conditions, thereby enabling adaptive and context-aware authentication decisions. By replacing rigid static-threshold mechanisms with LLM-driven contextual reasoning, the framework enhances authentication robustness under dynamic operating environments and behavioral variability. These capabilities create a new paradigm for behavioral biometric authentication in which authentication decisions are no longer based solely on static similarity matching but instead involve intelligent reasoning over behavioral signals and contextual information.
As a result, the proposed framework dynamically adapts to diverse behavioral and environmental conditions, improving authentication robustness under real-world variability. By replacing rigid static-threshold mechanisms with LLM-driven contextual reasoning, the framework establishes a new paradigm for behavioral biometric authentication in which decisions are guided not only by similarity matching but also by intelligent reasoning over behavioral and contextual information.

To evaluate the effectiveness of the proposed framework, %we employ a phone blowing-based behavioral biometric authentication system using 
we collected blow-acoustic signals from 50 participants under diverse contextual conditions involving variations in environmental settings (e.g., in quiet rooms and noisy public environments), postural modes (e.g., in sitting and standing positions), and phone interaction dynamics with different blowing patterns. The collected signals are processed using %conventional biometric
advanced feature extraction techniques, such as Gammatone Frequency Cepstral Coefficients (GFCCs) \cite{gfcc}, while the LLM reasoning engine analyzes contextual information and behavioral consistency to support adaptive authentication decisions. Experimental results demonstrate that the proposed LLM-assisted framework achieves an authentication accuracy of 99.46\% and 99.38\% under rule-based and score-based modes, respectively, both substantially outperforming the 99.09\% achieved by the conventional static threshold-based approach. 
%Experimental results demonstrate that the proposed LLM-assisted framework achieves an authentication accuracy of 99.46\%, compared to 99.09\% achieved by the conventional static threshold-based approach. 
More importantly, the proposed LLM-based framework significantly improves authentication robustness against behavioral variability, dynamic environmental conditions and adversarial contexts compared to conventional static threshold approaches.

%The proposed LLM-assisted framework improves security by introducing semantic and contextual reasoning into behavioral biometric authentication.
In general, by systematically integrating biometric similarity analysis with LLM-based contextual reasoning, the proposed approach transforms behavioral biometrics %authentication 
from a purely signal-level matching problem into a cognitive trust evaluation process capable of operating effectively under dynamic real-world conditions. Overall, this work makes the following main contributions:
\begin{itemize}
    %\item \optional{We identify fundamental limitations of conventional behavioral biometric authentication systems arising from static threshold-based decision mechanisms and limited contextual reasoning capability.}
    %\item We propose a novel LLM-assisted behavioral biometric authentication framework that integrates LLM's contextual reasoning capability into behavioral biometric authentication. % decision-making.
    %\item We develop an adaptive authentication strategy capable of dynamically interpreting behavioral variability and contextual changes using LLM reasoning.
    % \item We propose a novel LLM-assisted behavioral biometric authentication framework that integrates LLM-based contextual reasoning to enable adaptive authentication under dynamic behavioral variability and contextual changes.
    % \item We evaluate the proposed framework using on a novel phone blowing-based behavioral biometric technique exhibiting diverse behavioral and contextual variations. 
    % \item The experimental evaluations conducted on a dataset collected from 50 participants demonstrate that the proposed approach achieves an authentication accuracy of 99.43\%, outperforming its non-LLM counterpart while also significantly improving robustness against 
    % dynamic real-world conditions. 
    \item We propose a novel LLM-assisted behavioral biometric authentication framework that systematically integrates LLM-based contextual reasoning with %a novel blow-acoustic 
    behavioral biometric techniques to enable adaptive authentication under dynamic behavioral and contextual variations.
    \item We evaluate the proposed framework using a novel phone-blowing-based behavioral biometric technique that exhibits diverse behavioral and contextual variations.
    %\item Experimental results on a dataset collected from 50 participants demonstrate that the proposed approach achieves 99.46\% authentication accuracy, outperforming its non-LLM counterpart while significantly improving authentication robustness under dynamic real-world conditions.
    \item Experimental results on a dataset collected from 50 participants demonstrate that the proposed approach achieves 99.46\% and 99.38\% authentication accuracy under rule-based and score-based modes, respectively, outperforming its non-LLM counterpart while significantly improving authentication robustness under dynamic real-world conditions.
\end{itemize}

%%Despite the remarkable advances of Large Language Models (LLMs) in contextual understanding, sequential reasoning, multimodal interpretation, and adaptive decision-making, their application to behavioral biometric authentication remains largely unexplored. Existing behavioral biometric systems primarily focus on signal-level feature matching and static threshold-based verification, with limited capability to reason about contextual information, behavioral variability, or session-level interaction patterns. The powerful reasoning capabilities of LLMs create a significant opportunity to transform behavioral biometric authentication into a context-aware and cognitively driven security framework capable of adaptive trust evaluation, semantic behavioral understanding, and robust authentication under dynamic real-world conditions.

%\input{03_related_work}
%\input{04_background}
\section{Threat Model and Security Goals}\label{threat_model}

\subsection{Threat Model}

%In our threat model, we consider an adversary that targets behavioral biometric authentication systems. 
In behavioral biometric (or general) authentication systems, the primary objective of the adversary is to impersonate the legitimate user and gain unauthorized access to the protected system by bypassing the authentication system. Therefore, in our threat model, %we assume that the adversary may: %has the following capabilities:
we consider different attacks that commonly target behavioral biometric systems, which includes:
\begin{itemize}
    % \item Influence environmental conditions such as ambient acoustic noise.    
    % \item Induce changes in user interaction conditions.
    % \item Affect device orientation or phone handling dynamics.   
    % \item Cause behavioral variability during authentication sessions.
    \item \emph{Threshold-based attacks}: %Exploits weaknesses of static threshold-based authentication systems by generating 
    Generates near-threshold behavioral patterns to bypass the authentication.
    \item \emph{Mimicry attacks}: Imitates user’s behavioral patterns.
    \item \emph{Template compromise attacks}: Compromises stored biometric templates.
    \item \emph{Replay attacks}: Replays previously captured behavioral signals. 
    \item \emph{Synthetic attacks}: Generates artificial data using AI models and inject it to the authentication system. 
    %\item \emph{Context manipulation attacks}: Manipulates environmental or contextual conditions.
    \item  \emph{Model-based attacks}: Targets trained LLM models through techniques such as %model poisoning, 
    prompt injection and jailbreaking.
\end{itemize}

\subsection{Security Goals}

The proposed LLM-assisted behavioral biometric authentication framework focuses on improving authentication robustness under varying environmental, behavioral, and contextual conditions. %The primary objective is not to detect or prevent attacks involving biometric template theft, replay attacks, or precise behavioral imitation. Instead, the proposed framework 
We aim to address the limitations of conventional static threshold-based behavioral biometric techniques that often fail under dynamic real-world conditions due to natural behavioral variability and contextual changes. Therefore, the main objectives of this work are to achieve the following aspects. 
\begin{itemize}
    \item \emph{Accuracy}: As a behavioral biometric technique, we aim to achieve high accuracy to effectively minimize the risk of impersonation attacks (i.e., by minimizing false positives), while also enhancing user convenience by lowering the number of required authentication attempts (i.e., by minimizing false negatives). This requires accurately distinguishing legitimate users even under dynamic contextual conditions. 
    \item \emph{Resilience against known attacks}: We aim to achieve high resilience against the common behavioral biometric attacks mentioned above.
    \item \emph{Usability}: Usability is a primary challenge in behavioral biometrics. In this work, we aim to achieve the key usability issues, including:
    \begin{enumerate}
        \item \emph{User convenience}: We aim to achieve system's intuitiveness (easy to use) and non-intrusiveness (not causing discomfort).
        \item \emph{Contextual robustness}: We aim to achieve effectiveness of the system under different real-world conditions and contexts, including in different postural modes (sitting and standing), environmental conditions (quiet and noisy), %health status (sick or normal), 
        blowing intensity (low and high), and blowing patterns (simple and complex).  
    \end{enumerate}
\end{itemize}
\section{LABBA: Proposed Work}\label{proposed_work}
%\howard{the authors highlight that contextual information is incorporated into the authentication process. However, it is not clearly explained what exactly is meant by "contextual information" in this work or how the LLM contributes to incorporating or utilizing such information. A clearer explanation of these aspects is needed.

%We need to rewrite the spectral feature extraction section to either cite blowlive for the value used (50ms , 20ms, 13 coeff, etc). Or remove the specific value mentioned.
%} 

\begin{wrapfigure}{r}{0.42\textwidth}
    \vspace{-0.8cm}
    \centering  
    \includegraphics[width=0.32\textwidth]{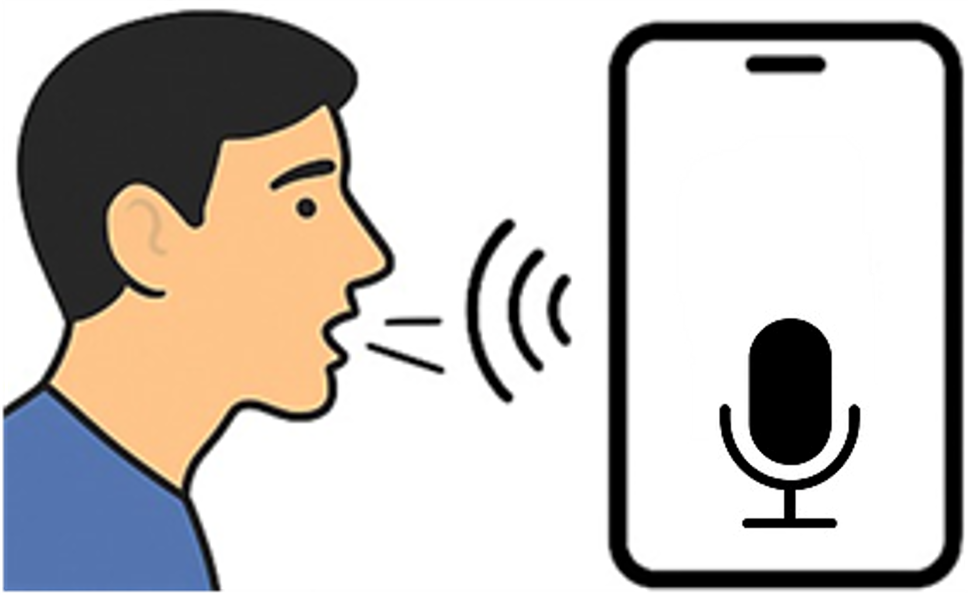}  %%%%0.35
    %\vspace{-0.6cm}
    \caption{Phone blowing illustration}
    \label{fig:phone_blowing_illustration}
    %\vspace{-0.3cm}
\end{wrapfigure}

This section presents the proposed LLM-assisted behavioral biometric authentication (LABBA) framework designed to enhance authentication robustness under dynamic contextual conditions. The framework integrates %conventional 
a blow-acoustic based behavioral biometric feature extraction 
%authentication technique 
process with LLM-based contextual reasoning to enable adaptive and context-aware authentication decision-making. In brief, the user first blows toward the phone screen (as illustrated in Figure \ref{fig:phone_blowing_illustration}) where blow-acoustic signals are captured via the phone's microphone. The captured signals consist of rich spectral and temporal features. %\eyasu{The following sentences are more technical, hence it would be better to move it to the workflow or other relevant sections. As an intro the the proposed work, this section should provide only a high-level information.}
%Initially, 
Spectral features are extracted in the form of Gammatone Frequency Cepstral Coefficients (GFCCs) \cite{gfcc} and subsequently passed to a trained Convolutional Neural Network (CNN) model to generate discriminative embedding representations. Concurrently, temporal features are extracted using a segmentation-based algorithm to capture user-specific blow interaction patterns. %\addnew{
The resulting spectral and temporal comparisons between the enrolled and query samples are then combined into a unified biometric representation, which is provided to a lightweight LLM for contextual evaluation and authentication reasoning. %} %The resulting spectral similarity measurements and temporal behavioral features are then aggregated into contextual information summaries, which are provided to a lightweight LLM for contextual evaluation and authentication reasoning. 
%Unlike traditional behavioral biometric systems that rely on static thresholds and fixed similarity-based verification, the proposed approach dynamically interprets behavioral biometric patterns alongside contextual information to improve resilience against behavioral variability and adversarial manipulation.
Unlike traditional behavioral biometric systems that rely on static thresholds and fixed similarity-based verification, the proposed approach dynamically interprets behavioral biometric patterns alongside contextual information to improve resilience against behavioral variability and adversarial manipulation. Here, contextual information refers to the user's enrolled interaction history, the temporal blow patterns of the query, and its session-level consistency with the user's own baseline, expressed through the spectral and temporal comparisons between the query and enrolled samples. The LLM reasons over this contextual information and returns an accept or reject decision together with a justification, while environmental conditions are reflected only through these comparisons and are not provided as separate inputs.

% \begin{figure}[htbp]
%     \centering  
%     %\vspace{-0.3cm}
%     \includegraphics[width=0.31\textwidth]{images/LABBA-Figure.png}  
%     \caption{Phone blowing illustration} %\eyasu{We need to remove the face image inside the phone if we don't consider the facial modality.}
%     \label{fig:phone_blowing_illustration}
%     %\vspace{-0.6cm}
% \end{figure}
%\begin{comment}
%\subsection{System Architecture}
%\begin{figure}[!htbp]
%\end{figure}
%\vspace{2cm}
%\end{comment}

%%\eyasu{How does the proposed approach achieve dynamically interprets behavioral biometric patterns and contextual information to make reasoning based authentication decision? Need to highlight the approach here.}
%\howard{okay}

\subsection{Workflow}

%The proposed framework follows the following pipelines (to be illustrated in figure):\\
%User Blow Interaction --> Blow-Acoustic Data Acquisition --> Data Preprocessing --> Feature Extraction --> Similarity Score Computation --> Context Collection Module (Environment, Session, Behavioral Metadata) --> LLM-Based Reasoning Engine --> Adaptive Authentication Decision
\begin{figure*}[htbp]
    \centering  
     \includegraphics[width=\textwidth]{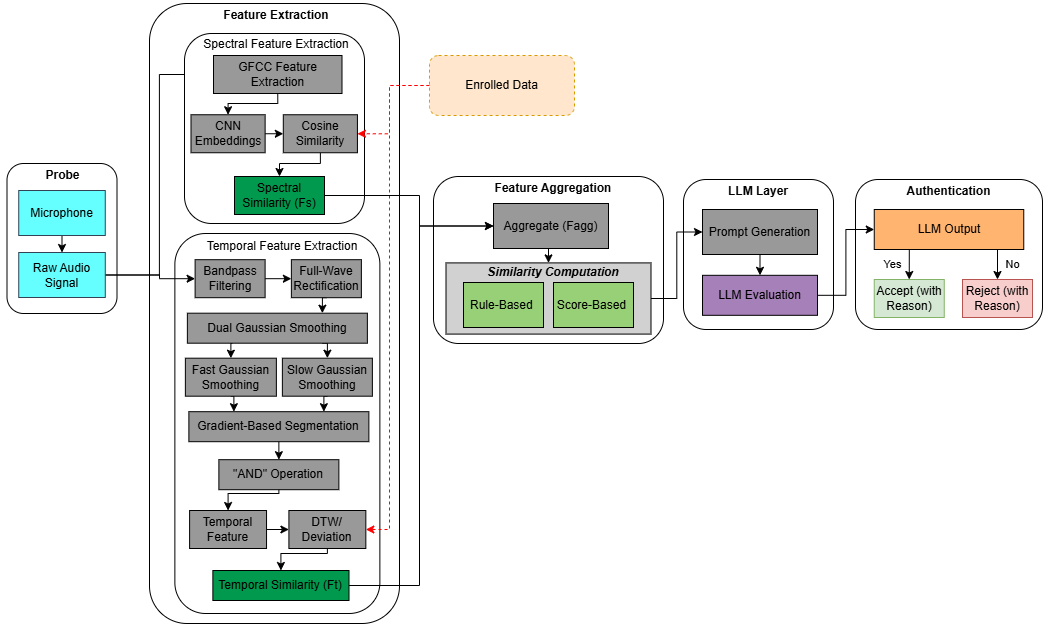} 
     %\vspace{-0.2cm}
    \caption{A high-level workflow of LABBA - \textit{Similarity Computation} uses either
Rule-Based or Score-Based, not both. Red dotted arrows indicate the enrolled-data comparison paths.}
    % \howard{will fix the image again. Just a sketch for the new one}
    \label{fig_LABBA_architecture}
\end{figure*}

%\howard{Additional discussion of what "enrolled data" is required here.}
Figure \ref{fig_LABBA_architecture} illustrates the overall workflow of LABBA. The framework operates through a multi-stage pipeline that integrates data collection, feature extraction, feature aggregation, and LLM-assisted evaluation to perform biometric authentication with contextual reasoning capabilities. Initially, raw %audio 
acoustic signals are collected by the phone's microphone %through users' blow attempts. 
as users blow toward the phone screen, as illustrated in Figure \ref{fig:phone_blowing_illustration}. The acquired recordings are subsequently processed using two complementary feature extraction mechanisms, namely spectral feature extraction and temporal feature extraction. Spectral feature extraction captures frequency-domain characteristics of the blow-acoustic signals, while temporal feature extraction models the structural and dynamic behavioral patterns present over time. The features of each attempt are then compared against the user's enrolled data, which consists of blow samples recorded from the legitimate user during enrollment. The resulting spectral and temporal comparisons are kept together for subsequent processing. %The extracted features are then aggregated into a unified representation for subsequent processing. 
Following feature aggregation, a prompt generation module constructs structured prompts tailored to the target LLM model by incorporating the unified feature representation. The generated prompts are subsequently utilized for LLM-assisted authentication and reasoning-based evaluation.

\subsection{Blow-Acoustics Data Acquisition} \label{data_acquisition}

The data acquisition stage focuses on capturing airflow-induced acoustic signals generated during phone-blowing interactions. In the proposed system, users perform blowing gestures toward the smartphone microphone screen, producing acoustic patterns that contain user-specific temporal and spectral characteristics suitable for behavioral biometric authentication. The phone-blowing interaction is performed within a 5-second duration, during which users may exhibit personalized blowing behaviors, including variations in the number of blows, duration of each blow, blowing intensity, blowing patterns, and airflow dynamics.

%During authentication, users blow air toward the smartphone microphone, generating airflow-induced acoustic signals that contain personalized temporal and spectral characteristics. Specificaly, blow-acoustic behavioral biometric signals were collected from 50 participants using smartphone microphones under diverse contextual conditions. The dataset includes variations involving environmental noise conditions, user interaction styles, number of blows per session, blowing intensity variations, session-level behavioral changes, ... Each participant performed multiple authentication sessions under varying real-world conditions to capture natural behavioral variability.

Blow-acoustic biometric signals provide several advantages, including strong resistance against spoofing attacks, implicit liveness detection, continuous and unobtrusive authentication capability, non-invasiveness, high usability, and low-cost deployment through existing smartphone microphones. %\cite{halim2025blowprint}. 
However, as a soft biometric modality, blow-acoustic signals are highly sensitive to contextual variability, including ambient acoustic noise, smartphone orientation, %\howard{from start to the end, we only mention sitting and standing. This one include others that we do not have. Should we align it or how?}
user posture (e.g., standing and sitting)%, sleeping, walking, and running)
, variations in blowing intensity, environmental dynamics, as well as user fatigue and behavioral drift. These challenges motivate the integration of LLM-based contextual understanding and logical reasoning into the authentication process to improve robustness under diverse real-world conditions.

\subsection{Feature Extraction}\label{feature_extraction}

The proposed framework extracts discriminative behavioral features from blow-acoustic signals to characterize user-specific airflow interaction patterns. The extracted features include two main categories: spectral and temporal. %following categories. 

\subsubsection{Spectral Feature Extraction} \label{spectral_feature_extraction} %\howard{there are many "magic numbers" in this section, mostly is from blowlive numbers that we reused. either cite the blowlive paper to show our choosing, or remove the mention of the numbers}
%\howard{this will be the same GFCC feature extraction explained in BlowLive Paper}\\
%\eyasu{Ok, please include any relevant information here.}
Spectral features are extracted from the blow-acoustic recordings using a GFCC-based processing pipeline to capture discriminative frequency characteristics and short-term temporal variations. We adopt the GFCC configuration of BlowLive~\cite{chekole2026blowlive}, the audio signal is segmented into overlapping frames of \emph{50 ms} with a \emph{20 ms} hop size. For each frame, a 24-channel gammatone filterbank is applied to obtain auditory-inspired spectral energy representations. The filterbank energies are then logarithmically compressed and transformed using a discrete cosine transform (DCT), where the first 13 coefficients are retained as the static GFCC representation: 
\begin{equation}\label{eq:gfcc_coefficient_vector}
\mathrm{g}(t_n) = \mathrm{DCT}\big(\log \mathrm{G}(t_n)\big),
\end{equation}
where $t_n$ denotes the $n$-th frame ($n = 1, \ldots, T$), $\mathbf{G}(t_n)$ is the vector of 24 gammatone filterbank channel energies at frame $t_n$, and $\mathbf{g}(t_n) = [g^{(1)}(t_n), \ldots, g^{(13)}(t_n)]^\top$ is the resulting static GFCC vector.

To further capture local shifts, first-order delta coefficients are computed using a symmetric regression window and concatenated with the static coefficients, resulting in a 26-dimensional feature vector for each frame. %\eyasu{Formal representation of the resulting feature vector?}
% \begin{equation}
% \mathrm{x}(t_n)=
% [g^{(1)}(t_n), \dots, g^{(13)}(t_n),
% \Delta g^{(1)}(t_n), \dots, \Delta g^{(13)}(t_n)]^\top
% \in \mathbb{R}^{26},
% \end{equation}
\begin{equation}
    \begin{aligned}
        \mathrm{x}(t_n) & = [g^{(1)}(t_n), \dots, g^{(13)}(t_n), \Delta g^{(1)}(t_n), \dots, \\
        & \Delta g^{(13)}(t_n)]^\top \in \mathbb{R}^{26},
    \end{aligned}
\end{equation}
where $\mathbf{x}(t_n)$ denotes the 26-dimensional feature vector at frame $t_n$ and
$\Delta g^{(k)}(t_n)$ denotes the corresponding first-order temporal derivative (delta coefficient).

To obtain stable and discriminative embeddings that are independent of temporal sequence length, the extracted GFCC sequences are subsequently processed using a CNN-based feature encoder. %\addnew{
The encoder was trained on 70\% of the sessions and evaluated on the remaining 30\%, so that test sessions are unseen during training. It was optimized with a triplet loss, which pulls embeddings of the same user's sessions together and pushes those of different users apart. %} 
The model consists of stacked 1D convolutional layers with batch normalization and Rectified Linear Unit (ReLU) \cite{relu} activation, followed by temporal mean pooling and fully connected layers. The final output is L2-normalized to produce a compact 128-dimensional spectral embedding representation for each blow-acoustic sample. %\eyasu{Formal representation of the resulting embedding?}
\begin{equation} \label{eq:z_embeddings}
\mathrm{z} = f_{\theta}(\mathrm{X}) \in \mathbb{R}^{128},
\end{equation}
where $\mathrm{X}\in\mathbb{R}^{T\times26}$ denotes the sequence of frame-level GFCC feature vectors over $T$ temporal frames, and $f_{\theta}:\mathbb{R}^{T\times26}\rightarrow\mathbb{R}^{128}$ represents the CNN-based feature encoder that maps the variable-length input sequence into a fixed-dimensional embedding space.

The extracted spectral embeddings are compared against enrolled user templates using cosine similarity. Suppose $\mathrm{E}$ and $\mathrm{Q}$ represent the embeddings $\mathbf{z}$ (Eq.~\ref{eq:z_embeddings}) extracted from the enrolled and query blow-acoustic signals, respectively. The cosine similarity score between the two is computed as:
\begin{equation}
F_{\mathrm{s}}(\mathrm{E, Q}) =
\frac{\sum_{i=1}^{n} e_i q_i}
{\sqrt{\sum_{i=1}^{n} e_i^2}\sqrt{\sum_{i=1}^{n} q_i^2}},
\end{equation}

where $\mathrm{E} = [e_1, e_2, ..., e_n]$ is the enrolled blow-acoustic embedding, $\mathrm{Q} = [q_1, q_2, ..., q_n]$ is the query blow-acoustic embedding, $n$ is the embedding dimension, and $F_{\mathrm{s}}$ denotes the resulting cosine similarity for the current authentication attempt.

In the conventional approach, this similarity score alone is used to make authentication decision. However, %the resulting similarity 
this score alone is insufficient for robust authentication under dynamic contexts and conditions discussed above, which motivates %us the 
our use of LLM-based contextual reasoning using temporal features, as discussed below.

%\eyasu{Please show how the resulting embedding is used in the final LLM decision making. If I am not wrong, the GFCC embeddings are used only for computing the biometric similarity score. If so, it needs to be shown clearly (possibly using a formal representation via equation). For example, the similarity computation content can be merged here. Then, it should be shown how it is used in the final LLM decision making. }\\ \howard{okay, I have moved the simiarity score computation here and align the equation to be used in the next section}\\ \eyasu{Nice. Please double-check consistency of the flow so that it is clear for readers.}

\subsubsection{Temporal Feature Extraction} \label{temporal_feature_extraction}
%\eyasu{This discussion should be supported by graphical illustration. For example, you can use the figure of a single session and indicate different regions of the figure, such as  inter-blow gaps, blow duration, average amplitude, number of blows, etc, and discuss all aspects in a clear way. Otherwise, reviewers may not understand what this section is talking about. } \\ \howard{okay, I have made the update. Can let me know if it is still too vague}\\ \eyasu{Looks good now. Will check it in details again. BTW, this one is an important section of the paper as the main contribution of this work is based on the extracted temporal features, which are used as contextual inputs to the LLM reasoning.}

%\howard{Do we move this figure to be a little higher? or we move the figure 3 down? because figure 3 is mentioned later than figure 4. Or is it okay?} \\ \eyasu{In that case, we can swap their placement.}

\begin{figure}[t]
     \centering
     \begin{subfigure}{0.4\textwidth}
         \centering
         \includegraphics[width=\linewidth]{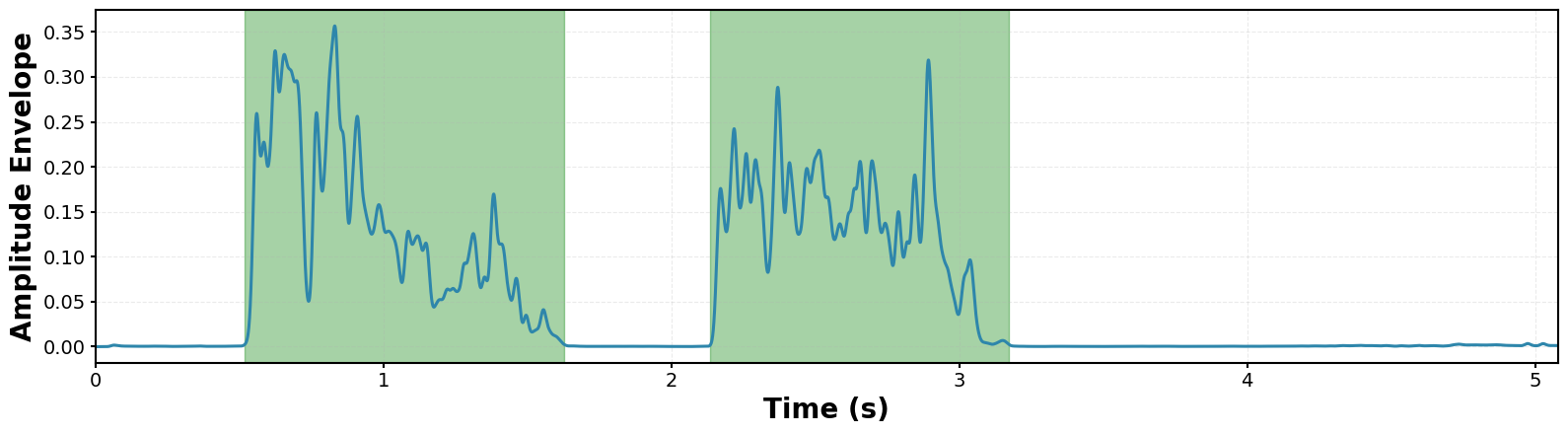}
         \caption{}
         \label{fig:fast_detector}
     \end{subfigure}
     \begin{subfigure}{0.4\textwidth}
         \centering
         \includegraphics[width=\linewidth]{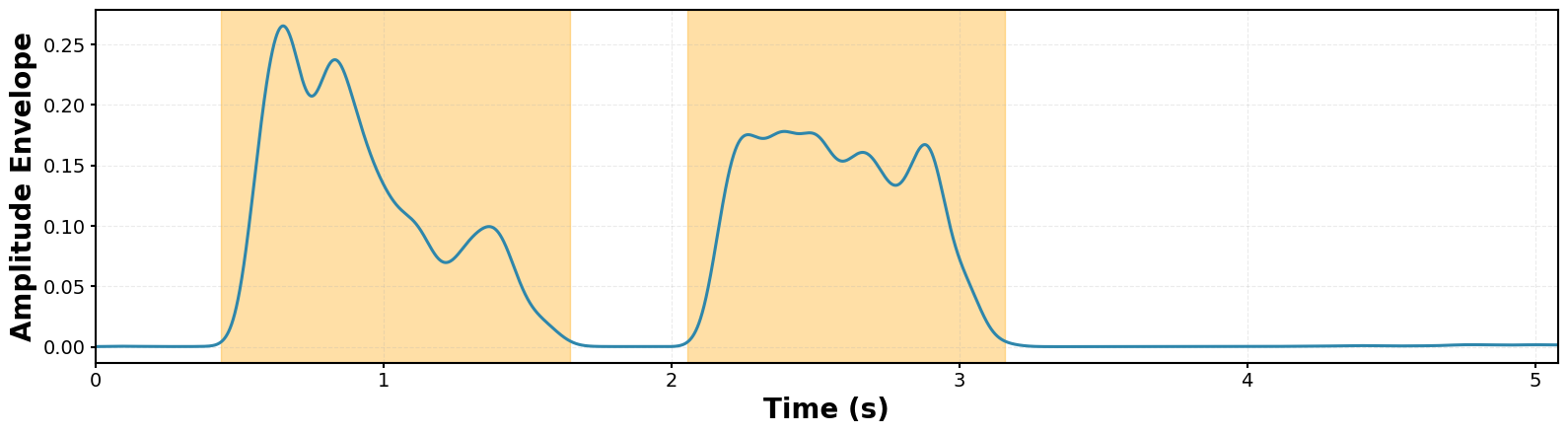}
         \caption{}
         \label{fig:slow_detector}
     \end{subfigure}
     \begin{subfigure}{0.4\textwidth}
         \centering
         \includegraphics[width=\linewidth]{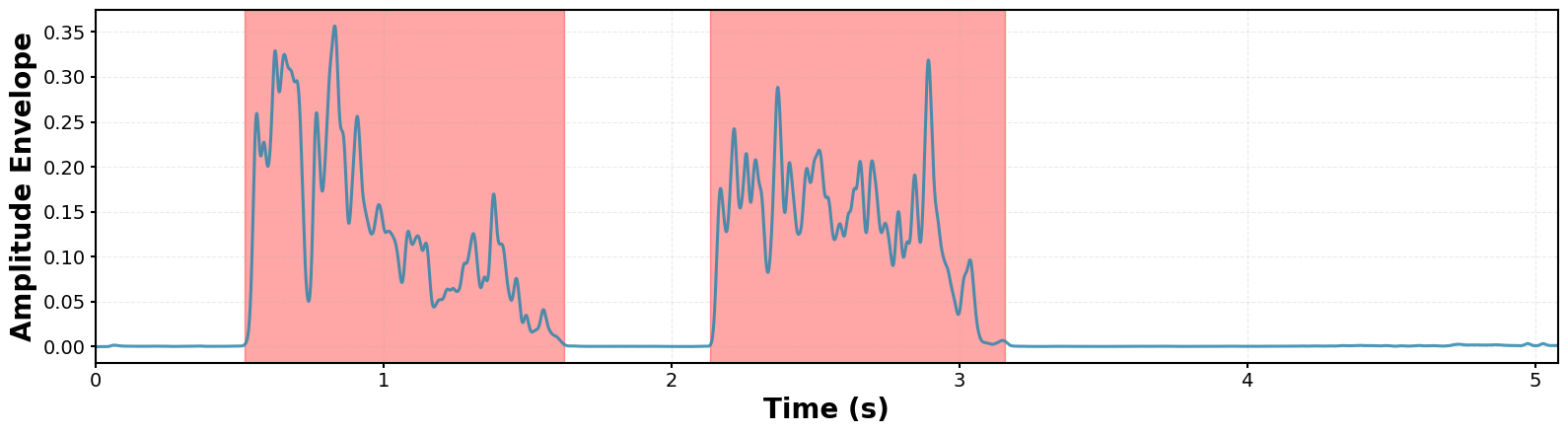}
         \caption{}
         \label{fig:fused_detector}
     \end{subfigure}
     \caption{Gaussian smoothing modes: (a) fast (b) slow (c) fused}
     \label{fig:guassian_detector}
\end{figure}

%\howard{the gaussian smoothing mentioned twice. I did add it on the later section. The figure mention twice. I have marked them both as revisit. Which place of explanation do you think it will be better at?} \\ \eyasu{Referring the figures twice is not a problem. But, please check the flow and decide where to put the gaussian smoothing discussion.}\\ \howard{I think early is better, I can remove the later one. But for the final one, it is better at the later portion as we discussed it as a AND combination.} \\ \eyasu{Sounds good.}
Temporal and blow-dynamic features are extracted to characterize user-specific blowing patterns. Given a 5-second blow-acoustic recording, the signal first undergoes bandpass filtering to suppress unwanted frequency components. The filtered signal is then full-wave rectified to obtain its amplitude envelope representation. To improve robustness against noise and intra-user variability, a dual-smoothing strategy is employed using %both fast and slow Gaussian smoothing kernels. 
Gaussian smoothing kernels with different temporal scales. The fast Gaussian smoothing (illustrated in Figure \ref{fig:fast_detector}), obtained using a smaller standard deviation ($\sigma_f$), produces an envelope that preserves fine temporal transitions and captures transient blow onsets. In contrast, the slow Gaussian smoothing (illustrated in Figure \ref{fig:slow_detector}), obtained using a larger standard deviation ($\sigma_s$), yields a more stable envelope that captures long-term energy trends and sustained blow activity. A fused representation (illustrated in Figure \ref{fig:fused_detector}) combines the outputs of both Gaussian smoothing processes through mask intersection to retain consistently detected blow activity while suppressing spurious detections. Let $E$ denote the amplitude envelope of the preprocessed acoustic signal obtained after bandpass filtering and full-wave rectification. This process can be expressed as:
\begin{equation}
E_{\mathrm{fast}} = \mathrm{GS}_{\sigma_f}(E), \quad
E_{\mathrm{slow}} = \mathrm{GS}_{\sigma_s}(E),
\end{equation}
where $E_{\mathrm{fast}}$ and $E_{\mathrm{slow}}$ denote the fast- and slow-smoothed envelope representations, respectively, and $\mathrm{GS}_{\sigma}(\cdot)$ denotes Gaussian smoothing with standard deviation $\sigma$, implemented as a convolution between the signal and a Gaussian kernel.

For each smoothed representation, segmentation is performed using a dynamic baseline gradient-tracking algorithm \cite{sharma2024optimized}. The first-order gradient of the smoothed envelope is computed to detect rapid energy transitions associated with blow onset activity. Once activated, a segment remains active as long as either the gradient magnitude remains sufficiently high or the signal energy exceeds an adaptive baseline estimated from preceding signal regions. To suppress false activations caused by transient fluctuations, a silence-reset mechanism is applied: if the gradient remains near zero and the signal energy stays below the adaptive baseline for a sustained duration, the active segment is terminated and the detector state is reset. The segmentation operator $\text{Seg}(\cdot)$ converts a smoothed continuous envelope into a binary sequence indicating active (blow) and inactive (non-blow) regions:
\begin{equation}
S_{\mathrm{fast}} = \text{Seg}(E_{\mathrm{fast}}), \quad
S_{\mathrm{slow}} = \text{Seg}(E_{\mathrm{slow}}),
\end{equation}
where $S_{\mathrm{fast}}$ and $S_{\mathrm{slow}}$ denote the binary segmentation sequences obtained from the fast and slow smoothed envelopes, respectively.

Following the segmentation stage, a post-processing refinement is applied to improve temporal consistency. Specifically, short inactive gaps between adjacent active regions are filled using a minimum-gap constraint, while segments shorter than a predefined duration threshold are removed. The above segmentation and post-processing pipeline is independently applied to both fast and slow smoothed representations, producing two candidate activity masks. %\howard{we have mention the figure 3c above. Do you think its needed to show it again here? or just remove illustrated description}
The fused blow regions are obtained by intersecting both %masks 
fast and slow smoothed representations using a logical AND operation (shown in Figure \ref{fig:fused_detector}), which reduces false positives while retaining consistently detected blow activity across both temporal resolutions. The fused blow regions ($T_f$) are defined as:
\begin{equation}
T_f = S_{\mathrm{fast}} \cap S_{\mathrm{slow}}.
\end{equation}

\begin{figure}[htb]
     \centering
     \includegraphics[width=0.9\linewidth]{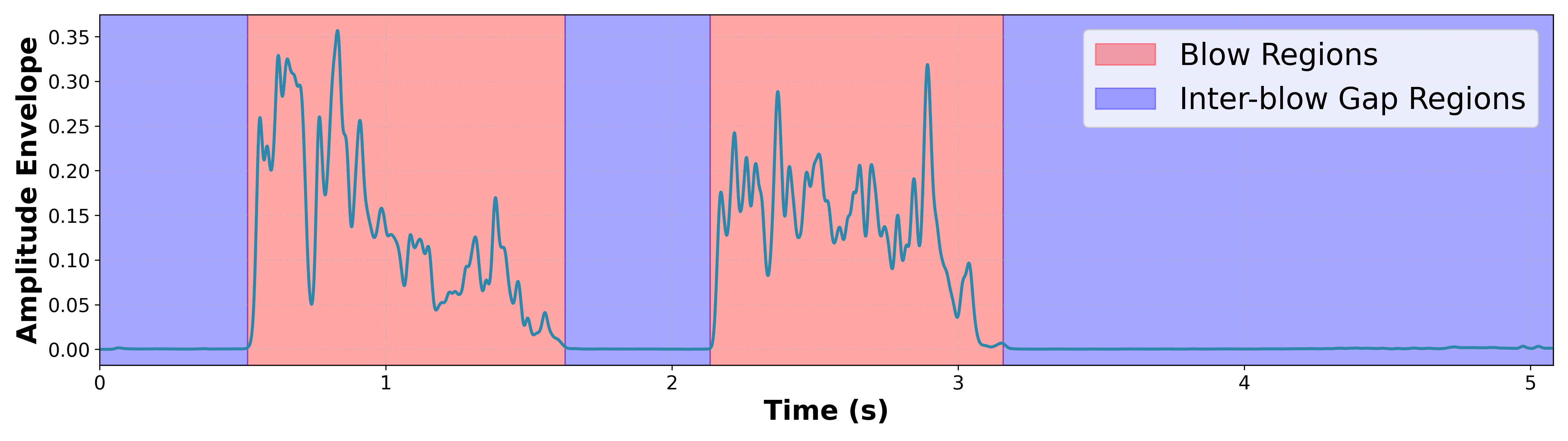}
     \caption{Temporal feature extraction}
     \label{fig:temporal_feature}
\end{figure}

The fused blow regions ($T_f$) are used to extract %\addnew{
the temporal feature set for the recording. %} %temporal features, including number of blows, inter-blow gaps, blow duration, and average amplitude, to represent user-specific behavioral patterns. 
Figure~\ref{fig:temporal_feature} illustrates the segmentation result for a single recording session, where the detected blow regions and inter-blow gaps are explicitly highlighted. Each contiguous blow region corresponds to an active blowing event, while the intervening non-active intervals represent inter-blow gaps%\addnew{
, including the boundary intervals at the beginning and end of the recording.
%}. %Based on this segmentation, temporal features are derived directly from the detected regions. 
%\addnew{
From $T_f$%}
, the number of blows ($N_b$) is obtained by counting the distinct blow regions, the duration of each blow ($D_b$) is computed from the temporal extent of its corresponding segment, and %\addnew{
the inter-blow gaps ($G_b$) correspond to the non-active intervals between consecutive blow regions. %} %Inter-blow gaps correspond to the non-active intervals between consecutive blow regions, including the boundary intervals at the beginning and end of the recording. 
%\addnew{
In addition, the RMS amplitude within each blow region ($A_b$) is computed from the bandpass-filtered signal, the duty cycle ($M_b$) is computed as the proportion of the session duration occupied by blow activity, and the blow shape ($Sh_b$) is characterized by the rise and fall times around each blow's peak amplitude.%}
%In addition, the average amplitude within each blow region is computed by averaging the envelope magnitude, providing a measure of blowing intensity.

\begin{comment}
The temporal feature representation ($\mathrm{F}_{t}$) is formally defined as:
\[
\mathrm{F}_{t}\ = \left\{ N_b,\; D_b,\; G_b,\; A_b \right\},
\]
where $N_b$ denotes the number of detected blow regions, $D_b$ represents the blow durations, $G_b$ represents the inter-blow gaps, and $A_b$ denotes the average amplitudes computed from the detected blow regions.
\end{comment}
%\addnew{
The resulting per-session temporal feature representation ($\mathrm{T}$), derived from $T_f$, is formally defined as
\[
\mathrm{T} = \{N_b, D_b, G_b, A_b, M_b, Sh_b\},
\]
%where $N_b$ denotes the number of detected blow regions, $D_b$ represents the blow durations, $G_b$ represents the inter-blow gaps, $A_b$ denotes the RMS amplitudes computed from the detected blow regions, $M_b$ denotes the duty cycle (the proportion of the session occupied by blow activity), and $Sh_b$ denotes the blow shape, characterized by the rise and fall times around each blow's peak amplitude.
where $N_b$, $D_b$, $G_b$, $A_b$, $M_b$, and $Sh_b$ denote the blow count, blow durations, inter-blow gaps, RMS amplitudes, duty cycle, and blow shape described above, respectively.

Since $\mathrm{T}$ is extracted independently for each recording, it is not yet a comparison quantity. To obtain the temporal feature representation used in authentication, denoted $\mathrm{F}_t$, the enrolled and query sessions' temporal features, $\mathrm{T}^{enroll}$ and $\mathrm{T}^{attempt}$, are compared channel-wise: sequential channels ($D_b$, $G_b$, $Sh_b$) are compared using Dynamic Time Warping (DTW) alignment distance, while scalar channels ($N_b$, $A_b$, $M_b$) are compared using the absolute deviation between the enrolled and query values:
\[
\mathrm{F}_t = \Delta(\mathrm{T}^{enroll}, \mathrm{T}^{attempt}),
\]
where $\Delta(\cdot)$ denotes this per-channel comparison, and $\mathrm{F}_t$ denotes the resulting raw temporal deviation for the current authentication attempt, analogous in form to the raw spectral similarity $\mathrm{F}_s(E,Q)$ defined in Section~\ref{spectral_feature_extraction}. 

\subsection{Feature Aggregation} \label{feature_aggregation}%\howard{need to rechoose words used in this discussion. We do not combined into a single unified set. We use them together as input. No concatenation or anything happened. }
To enhance authentication robustness under varying behavioral and environmental conditions, the proposed framework incorporates both spectral and temporal features described in Section~\ref{feature_extraction} and integrates them with LLM-based contextual reasoning. This enables adaptive context-aware authentication through the joint analysis of % temporal behavioral characteristics and biometric similarity scores computed using cosine similarity.
%\addnew{
spectral and temporal biometric characteristics, rather than relying on a single similarity measure alone.%}

%\addnew{
%To facilitate LLM-based reasoning, the extracted spectral and temporal features are first combined through a feature aggregation stage, which takes both feature domains and produces a single unified set of biometric features describing the authentication attempt:

Since $\mathrm{F}_{s}$ and $\mathrm{F}_{t}$ are both already expressed as comparisons between the enrolled and query samples, they can be provided together as the input describing the authentication attempt:
%they can be directly combined into a single unified set of biometric features describing the authentication attempt:
\begin{equation}
\mathrm{F}_{agg} = (\mathrm{F}_{s}, \mathrm{F}_{t}),
\end{equation}
where $\mathrm{F}_{agg}$ denotes %the aggregated feature, representing a single, unified comparison outcome for the current authentication attempt.
the pair of comparison outputs for the current authentication attempt, each retained as a separate piece of evidence.

% Since lightweight LLMs generally exhibit limited capability in directly interpreting raw numerical biometric features, a structured intermediate representation is introduced. A similarity scoring function $\mathcal{R}(\cdot)$ is applied to the enrolled and attempted feature representations jointly, converting the deviation between the two into a comparable similarity score for each channel of $\mathrm{F}_{agg}$:
% \begin{equation}
% \mathrm{P} = \mathcal{R}(\mathrm{F}_{agg}^{enroll}, \mathrm{F}_{agg}^{attempt}),
% \end{equation}
% where $\mathrm{F}_{agg}^{enroll}$ and $\mathrm{F}_{agg}^{attempt}$ denote the aggregated feature representations of the enrolled and attempted authentication samples, respectively, and $\mathrm{P}$ denotes the resulting interpreted, unified representation.
% \begin{comment}
% Since lightweight LLMs generally exhibit limited capability in directly interpreting raw numerical biometric features, a structured intermediate representation is introduced. A similarity scoring function $\mathcal{R}(\cdot)$ normalizes each channel of $\mathrm{F}_{agg}$ against the enrolled user's own baseline reference statistics $\mathrm{U}_{b}$:
% \begin{equation}
% \mathrm{P} = \mathcal{R}(\mathrm{F}_{agg}, \mathrm{U}_{b}),
% \end{equation}
% where $\mathrm{P}$ denotes the resulting unified representation, consisting
% of normalized similarity scores (0-100).
% \end{comment}
Since lightweight LLMs generally exhibit limited capability in directly interpreting raw numerical biometric features, a structured intermediate representation is introduced. A similarity computation function $\mathcal{R}(\cdot)$ converts each channel of $\mathrm{F}_{agg}$ into a comparable similarity representation:
\begin{equation}
\mathrm{P} = \mathcal{R}(\mathrm{F}_{agg}),
\end{equation}
where $\mathrm{P}$ denotes the resulting unified similarity representation, consisting of normalized similarity scores. We evaluate two variants of $\mathcal{R}(\cdot)$: a \textit{rule-based}  mode, described in Section~\ref{sec:rule_based_scoring}, and a \textit{score-based} mode, described in Section~\ref{sec:score_based_scoring}.

\subsubsection{Rule-Based} \label{sec:rule_based_scoring}
The rule-based variant converts each channel deviation into a discrete similarity category \{\text{VERY\_GOOD}, \text{GOOD}, \text{OKAY}, \text{WEAK}, \text{BAD}\} using fixed, globally-defined percentage-difference thresholds between the enrolled and attempted values, applied uniformly across all users. The embedding similarity channel is instead labeled directly from the cosine-vs-threshold margin, divided into the same five categories based on its position relative to the user's decision threshold and the borderline band. Blow count is compared directly as an integer difference rather than a percentage, while the remaining channels (gap, duration, shape, amplitude, and mass) use symmetric percentage differences, with the shape channel calibrated to wider thresholds to reflect its larger typical variation. %The resulting categorical labels for all six channels constitute $\mathrm{P}$ under this mode.

\subsubsection{Score-Based} \label{sec:score_based_scoring}
The score-based variant instead normalizes each channel of $\mathrm{F}_{agg}$ against the enrolled user's own baseline reference statistics $\mathrm{U}_b$. By converting raw spectral and temporal deviations into a common, normalized scale, this feature aggregation stage allows the lightweight LLM to perform more effectively. Lightweight LLMs are generally less capable of directly interpreting raw numerical values. By expressing every feature channel as a bounded, uniformly interpretable similarity score, the aggregation stage removes this burden from the downstream reasoning module, allowing the LLM to focus on weighing and reasoning over the scores themselves rather than interpreting their raw numerical meaning.

\subsection{LLM-Based Contextual Reasoning Engine} \label{llm_based_context}
The proposed framework integrates a lightweight LLM as a high-level reasoning module for adaptive authentication. The LLM is responsible for interpreting structured behavioral representations and biometric similarity scores to perform context-aware authentication decisions that go beyond threshold-based verification. To enable effective reasoning, prompt engineering is employed to construct a structured and semantically consistent input prompt from the outputs of the feature aggregation module (cf. Section \ref{feature_aggregation}). %\addnew{
The designed prompt incorporates the %interpreted similarity scores 
unified similarity representation $\mathrm{P}$ derived from the similarity computation function $\mathcal{R}(\cdot)$, allowing the lightweight LLM to perform context-aware authentication reasoning.%}
%The designed prompt incorporates biometric similarity scores computed using cosine similarity together with symbolic behavioral representations derived from rule-based interpretation and symbolic encoding, allowing the lightweight LLM to perform context-aware authentication reasoning.

\begin{comment}

Given the symbolic behavioral representation $\mathrm{P}$, a prompt generation function $\mathcal{G}(\cdot)$ is applied to construct the final LLM input prompt:
\begin{equation}
\text{Prompt} = \mathcal{G}(\mathrm{P}),
\end{equation}
where $\text{Prompt}$ represents the prompt generated.
    
\end{comment}
%\addnew{
Given the %interpreted 
unified similarity representation $\mathrm{P}$, a prompt generation function $\mathcal{G}(\cdot)$ is applied to construct the final LLM input prompt:
\begin{equation}
\text{Prompt} = \mathcal{G}(\mathrm{P}),
\end{equation}
where $\text{Prompt}$ represents the prompt generated.

%}

The generated prompt $\text{Prompt}$ is subsequently processed by the lightweight LLM to perform contextual authentication reasoning:
\begin{equation}
\mathrm{O} = \mathcal{L}(\text{Prompt}),
\end{equation}
where $\mathcal{L}(\cdot)$ denotes the lightweight LLM reasoning function and $\mathrm{O}$ represents the generated authentication response.

The generated response $\mathrm{O}$ consists of the authentication decision $D \in \{\texttt{ACCEPT}, \texttt{REJECT}\}$ accompanied by a short natural language justification $J$ that %justifies the decision based on the provided biometric and behavioral evidence.
explains the reasoning behind the decision based on the provided biometric and behavioral evidence.

\subsection{Implementation Details}
\begin{comment}

The proposed LABBA framework was implemented as a multi-stage behavioral biometric authentication system integrating GFCC-based spectral feature extraction, temporal behavioral feature extraction, feature aggregation, and lightweight LLM-based contextual reasoning.

Firstly, blow-acoustic recordings are concurrently processed by two feature extraction modules, namely spectral feature extraction (Section~\ref{spectral_feature_extraction}) and temporal feature extraction (Section~\ref{temporal_feature_extraction}). The resulting outputs are subsequently combined using the feature aggregation module (Section~\ref{feature_aggregation}), where spectral similarity scores and temporal behavioral descriptors are jointly encoded into a symbolic representation.

This aggregated symbolic representation is then transformed into a structured prompt using prompt engineering and processed by the lightweight LLM-based contextual reasoning module (Section~\ref{llm_based_context}), which performs prompt interpretation and generates the final authentication response consisting of an authentication decision ($\texttt{ACCEPT}$ or $\texttt{REJECT}$) and a short natural language explanation that justifies the decision based on the provided biometric and behavioral evidence.

\end{comment}

The proposed LABBA framework was implemented as a multi-stage behavioral biometric authentication system using Python. To collect blow-acoustic data, we develop a custom blow-based biometric authentication application for the Android platform. The application captures blow-acoustic signals through the device microphone at a sampling rate of 48~kHz. During acquisition, the raw audio signal is continuously buffered internally for the entire recording duration. Each recording session lasts for 5 seconds and contains the complete temporal acoustic waveform generated during the user’s blowing interaction. Blow-acoustic spectral features were extracted using the \texttt{Gammatone}\cite{gammatone_pypi} Python library for GFCC-based processing and CNN models implemented in TensorFlow, while blow-acoustic temporal %behavioral 
features were derived using pure Python with NumPy and SciPy. %\addnew{
Feature aggregation was implemented using custom Python logic. Per-user baseline reference statistics ($\mathrm{U}_{b}$) are precomputed once from each user's enrollment sessions. For each authentication attempt, the enrolled and query samples' spectral and temporal features are compared to obtain $\mathrm{F}_{s}$ and $\mathrm{F}_{t}$. Under rule-based mode, the similarity computation module $\mathcal{R}(\cdot)$ maps $\mathrm{F}_s$ and $\mathrm{F}_t$ onto a fixed set of discrete categorical labels using globally-defined thresholds, requiring no per-user baseline precomputation. Under score-based mode, per-user baseline reference statistics ($\mathrm{U}_b$) are instead precomputed once from each user's enrollment sessions, and $\mathcal{R}(\cdot)$ normalizes $\mathrm{F}_s$ and $\mathrm{F}_t$ against the corresponding
user's precomputed baseline to produce a continuous, bounded representation. In both cases, the resulting unified similarity representation $\mathrm{P}$ is suitable for LLM-based reasoning and prompt construction.
% }
%Feature aggregation was implemented using custom Python logic in two stages: a rule-based interpretation module that processes spectral and temporal feature representations to derive intermediate structured descriptors, followed by a symbolic feature extraction step that produces a unified symbolic representation for LLM-based reasoning and prompt construction. 
The resulting aggregated representation was transformed into prompts using prompt engineering and provided to a lightweight LLM, specifically \textit{Phi-3 Mini Instruct} (Q4 quantized variant)\cite{phi3mini4kgguf2024}, integrated via the \texttt{llama-cpp-python}\cite{llama_cpp_python} library, which performed contextual reasoning and generated the final authentication decision.

The source code and artifacts of LABBA are available at \cite{labba_dataset}. 
\section{Evaluation}\label{evaluation}

\subsection{Experimental Setup}

% \revisit{add laptop specification to run the experiment}\\
%To conduct the experiment, we develop a custom blow-based biometric authentication application for the Android platform. The application captures blow-acoustic signals through the device microphone at a sampling rate of 48~kHz. During acquisition, the raw audio signal is continuously buffered internally for the entire recording duration. Each recording session lasts for 5 seconds and contains the complete temporal acoustic waveform generated during the user’s blowing interaction.
To collect blow-acoustic experimental data, we use the blow-acoustic application we developed. %The collected recordings were subsequently processed using the feature extraction pipeline presented in Section \ref{feature_extraction}. 
The extracted spectral and temporal blow-acoustic features were provided to a lightweight LLM model, specifically Phi-3 Mini Instruct (Q4 quantized version)\cite{phi3mini4kgguf2024}, to generate authentication reasoning based on the captured blow-acoustic characteristics. The system was implemented using python and executed entirely on CPU-based processing without GPU acceleration. All experiments were conducted on a Dell XPS 15 9510 laptop equipped with an 11th Gen Intel Core i7-11800H processor (8 cores) and 32 GB RAM, running Windows.

% \eyasu{The below content should be moved to the accuracy evaluation section.}\\ \howard{okay, will move it when writing the accuracy evaluation section}
% To evaluate the performance of LABBA, recordings originating from the same participant were treated as genuine attempts, while recordings from different participants were treated as adversary attempts. Pairwise comparisons were performed across sessions to evaluate intra-user consistency and inter-user separability under varying environmental conditions. System performance was evaluated using standard biometric authentication metrics, including Accuracy, False Acceptance Rate (FAR), and False Rejection Rate (FRR) %Equal Error Rate (EER).

\subsection{Dataset}

\begin{figure*}[htbp]
     \centering
     \begin{subfigure}{0.325\textwidth}
         \centering
         \includegraphics[width=\linewidth]{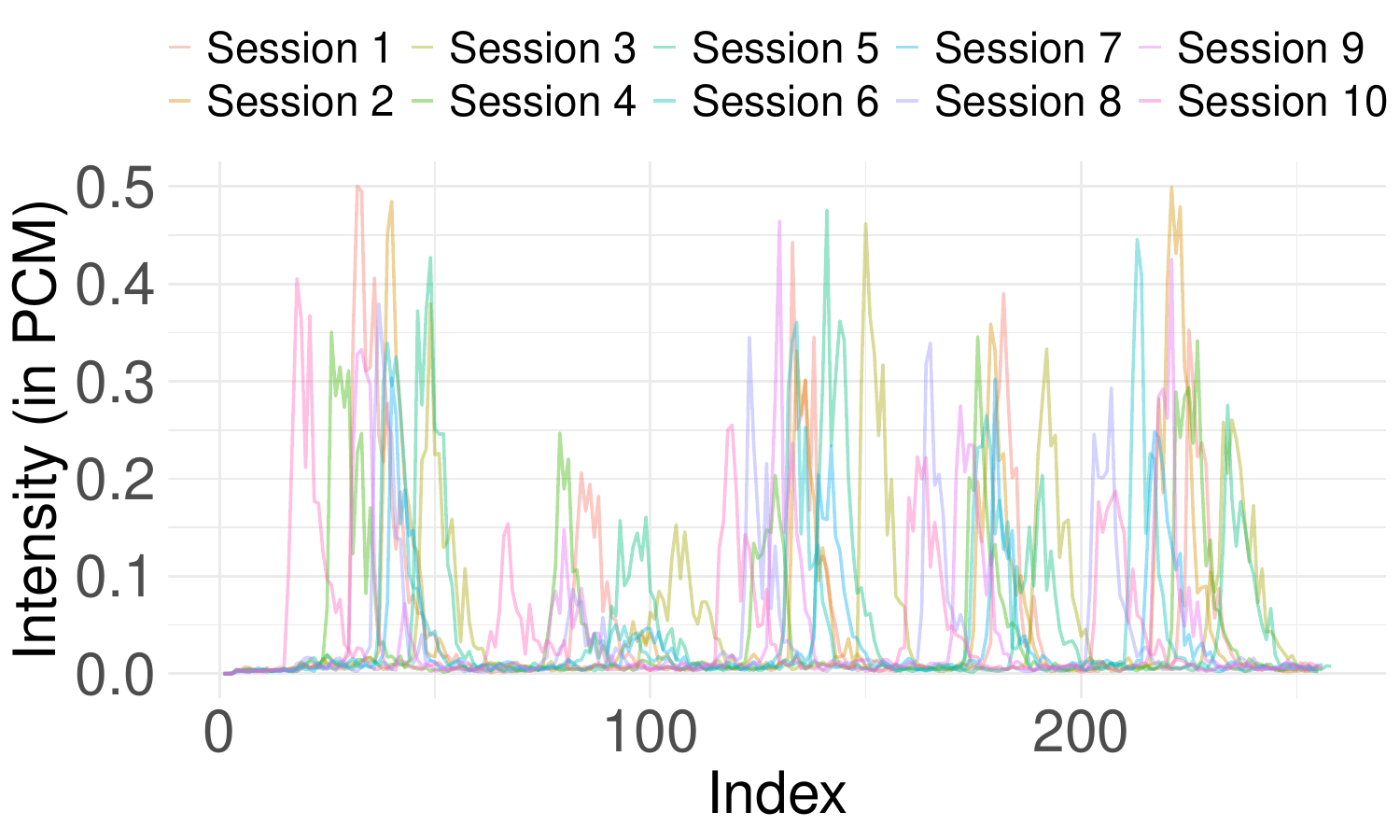}
         \caption{}
         \label{fig_participant1_raw}
     \end{subfigure}
     \begin{subfigure}{0.325\textwidth}
         \centering
         \includegraphics[width=\linewidth]{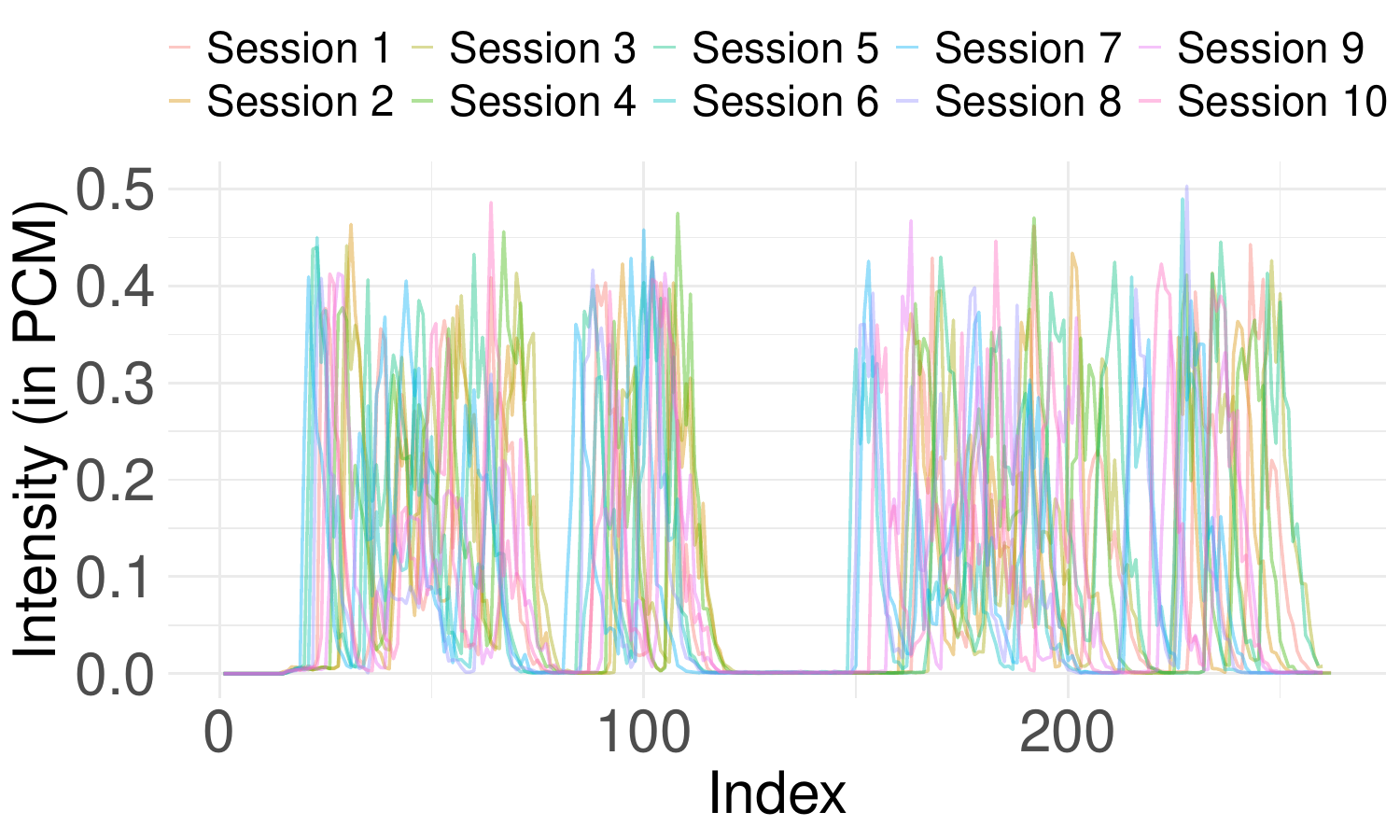}
         \caption{}
         \label{fig_participant2_raw}
     \end{subfigure}
      \begin{subfigure}{0.325\textwidth}
         \centering
         \includegraphics[width=\linewidth]{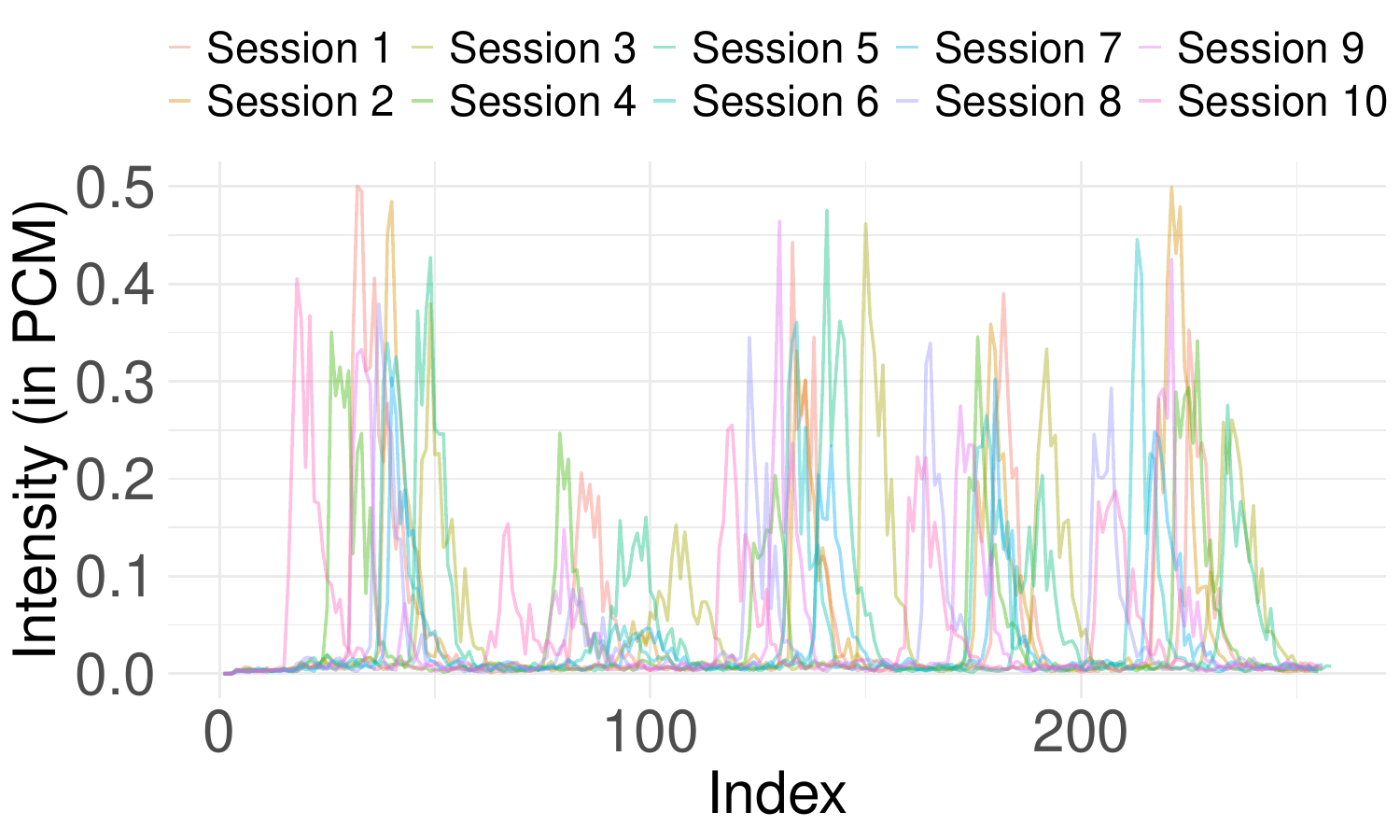}
         \caption{}
         \label{fig_participant3_raw}
     \end{subfigure}
     % \begin{subfigure}{0.49\textwidth}
     %     \centering
     %     \includegraphics[width=\linewidth]{images/participant_4.jpeg}
     %     \caption{}
     %     \label{fig_participant4}
     % \end{subfigure}
     %\vspace{-0.3cm}
     \caption{Sample raw blow-acoustic data of (a) Participant 1 (b) Participant 2 (c) Participant 3} %%%%, %(d) Participant 4, 
     %%%%where ``Signature'' is the DBA-based aggregated data of the 10 sessions}
     \label{fig:sample_raw_acoustic_data}
\end{figure*}

\begin{figure*}[htbp]
     \centering
     \begin{subfigure}{0.325\textwidth}
         \centering
         \includegraphics[width=\linewidth]{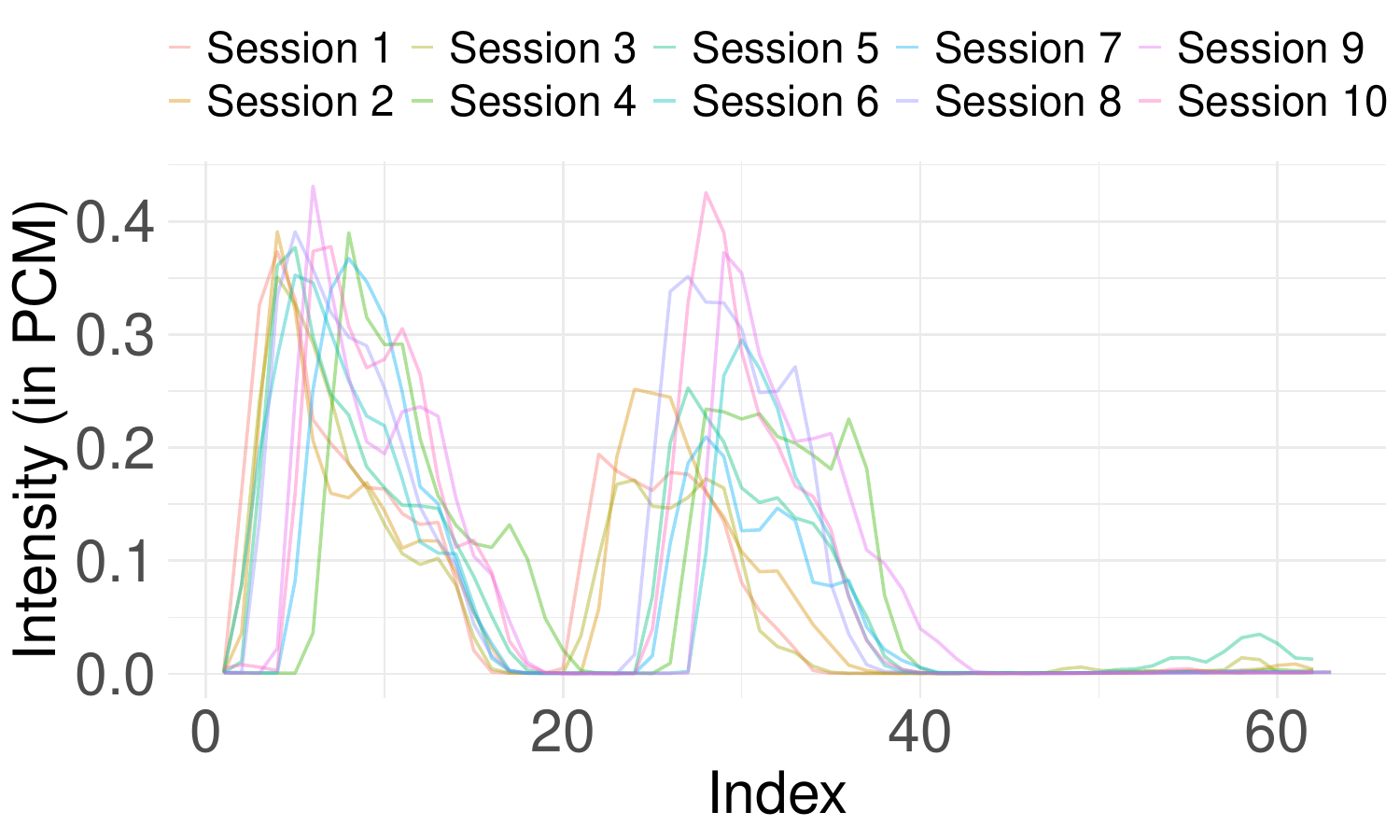}
         \caption{}
         \label{fig_participant1_refined}
     \end{subfigure}
     \begin{subfigure}{0.325\textwidth}
         \centering
         \includegraphics[width=\linewidth]{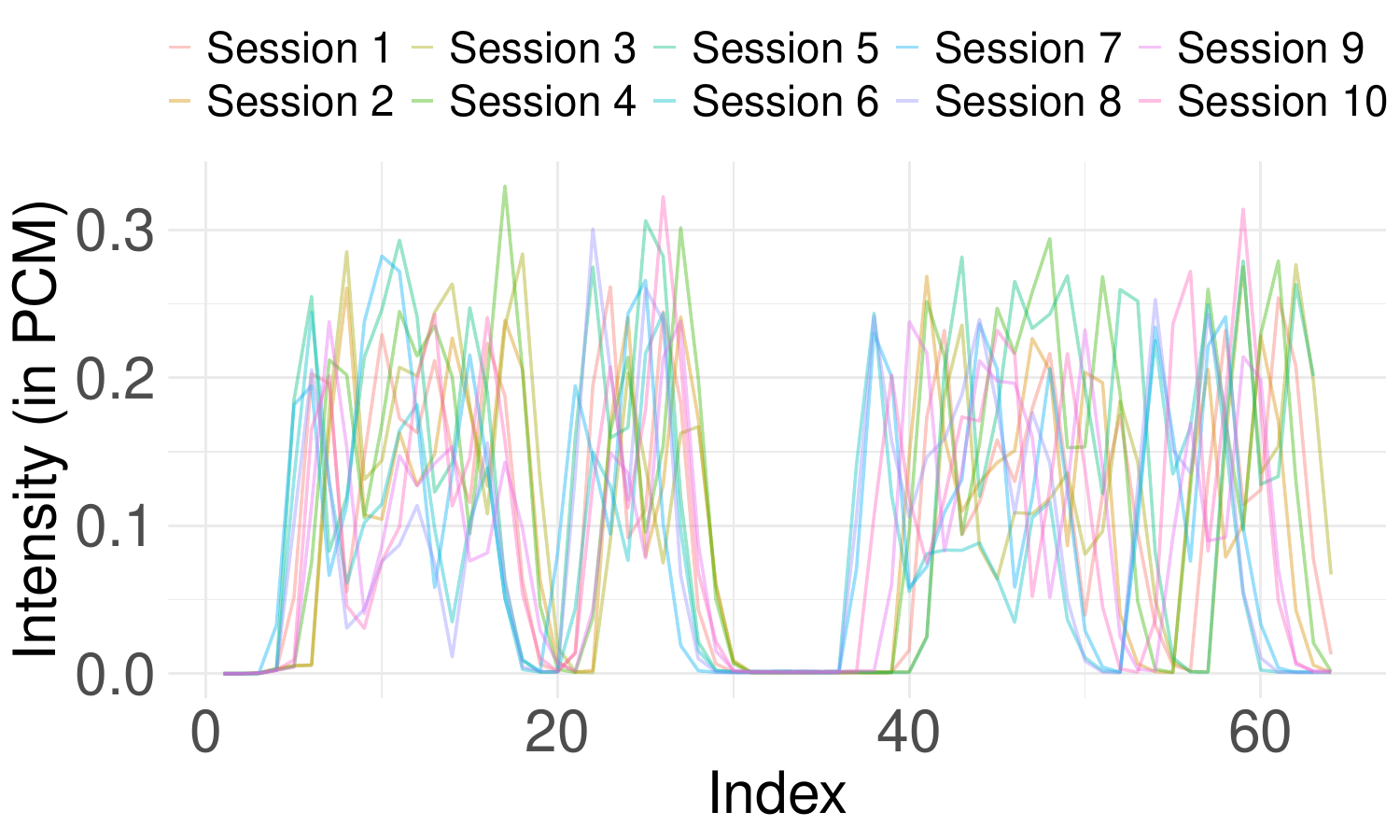}
         \caption{}
         \label{fig_participant2_refined}
     \end{subfigure}
      \begin{subfigure}{0.325\textwidth}
         \centering
         \includegraphics[width=\linewidth]{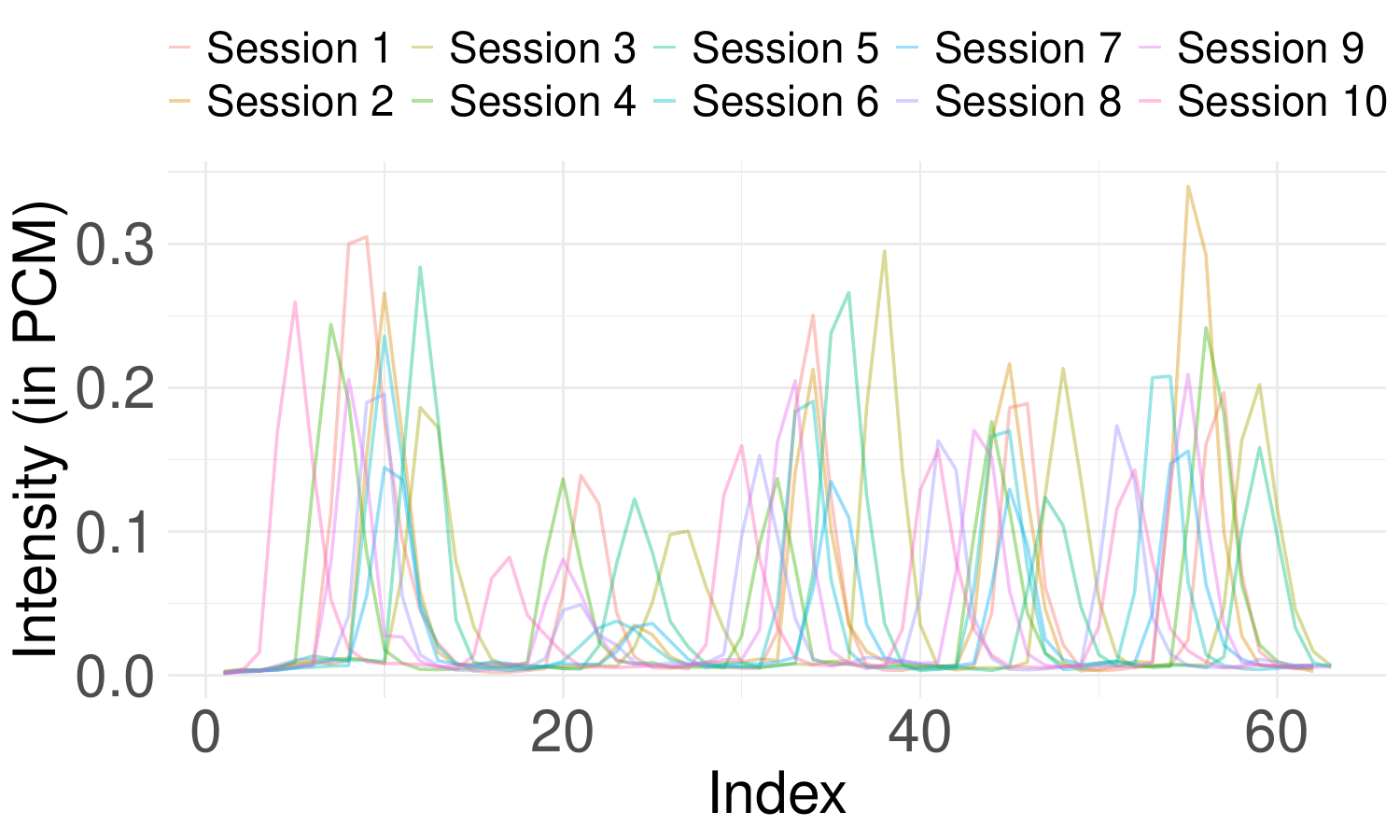}
         \caption{}
         \label{fig_participant3_refined}
     \end{subfigure}
     % \begin{subfigure}{0.49\textwidth}
     %     \centering
     %     \includegraphics[width=\linewidth]{images/participant_4.jpeg}
     %     \caption{}
     %     \label{fig_participant4}
     % \end{subfigure}
     %\vspace{-0.3cm}
     \caption{Sample refined blow-acoustic data of (a) Participant 1 (b) Participant 2 (c) Participant 3} %%%%, %(d) Participant 4, 
     %%%%where ``Signature'' is the DBA-based aggregated data of the 10 sessions}
     %\howard{let me know if this is good enough \\ \eyasu{Yes, good enough, thanks.}}
     \label{fig:sample_refined_acoustic_data}
\end{figure*}

We conducted a user study involving 50 participants (40 males, 10 females) randomly selected from demographic groups ranging from 22 to 65 years of age. Each participant performed 10 sessions, with each session lasting 5 seconds, resulting in 500 total attempts. To ensure robustness and evaluate real-world applicability, %the experiment was conducted in diverse environmental conditions, in quiet and noisy public environments. 
%\addnew{
the experiment was conducted under uncontrolled acoustic conditions, without a dedicated recording booth or acoustically treated room. %}
Each participant completed all their sessions in their own chosen environment%\addnew{
, reflecting the ambient acoustic variability typical of real-world smartphone usage, %} 
and they %different participants 
performed the experiment in both standing and sitting postural modes. % (standing and sitting). 
To show the patterns and characteristics of the collected dataset, we illustrate the raw and refined blow-acoustic data of three participants in Figure \ref{fig:sample_raw_acoustic_data} and \ref{fig:sample_refined_acoustic_data}, respectively. 

%\begin{comment}

%\end{comment}

\subsection{Accuracy Evaluation} \label{sec:accuracy_evaluation}
%\howard{Need an update of this section after the whole running process is done}
To evaluate the accuracy of LABBA, recordings originating from the same participant were treated as genuine attempts, while recordings from different participants were treated as adversary attempts. Pairwise comparisons were performed across sessions to evaluate intra-user consistency and inter-user separability under varying environmental conditions. Effectiveness of LABBA %System performance 
was evaluated using standard biometric authentication metrics, including Accuracy, False Acceptance Rate (FAR), False Rejection Rate (FRR), and Equal Error Rate (EER). The EER is commonly defined as the operating point at which the %False Acceptance Rate and False Rejection Rate 
FAR and FRR are equal. In this work, due to the discrete nature of the decision scores and the finite set of evaluation thresholds, we approximate the EER using the mean operating error rate, computed as:
%\begin{equation}
$\mathrm{EER} \approx \frac{\mathrm{FAR} + \mathrm{FRR}}{2}$.
%\end{equation}

% \multicolumn{2}{l}{Without LLM-Assisted}   & 0.9909 & 0.0013 & 0.4364 & 0.2189 \\ \midrule[0.1pt]
% \multirow{3}{*}{With LLM-Assisted} & Fast  & 0.9932 & 0.0021 & 0.2647 & 0.1334 \\
%                                    & Slow  & 0.9931 & 0.0023 & 0.2580 &  0.1302\\
%                                    & Fused & 0.9943 & 0.0023 & 0.1909 & 0.0966\\ \bottomrule
% \end{tabular}

Table~\ref{tab:evaluation} summarizes the biometric authentication accuracy of LABBA under configurations without and with LLM-assisted evaluation. The non-LLM baseline scheme achieves an accuracy of 99.09\%, FAR of 0.13\%, FRR of 43.64\% and EER of 21.89\%. 
%For the LLM-assisted scheme, accuracy is evaluated on three temporal segmentation configurations based upon the three Gaussian smoothing modes, namely Fast, Slow, and Fused. Among which, the Fused configuration achieves the best overall performance with an accuracy of 0.9943, FAR of 0.0023, FRR of 0.1909 and EER of 0.0966. %improving over the EER of 0.2189 achieved by the configuration without LLM-assisted evaluation. 
%Overall, %the results show that incorporating 
%the LLM-assisted scheme %achieves a significant achievement over 
%significantly outperforms the non-LLM scheme. %reasoning improves authentication performance compared to the non-LLM baseline. 
%$Overall, the results show that incorporating LLM-assisted reasoning leads to a slight improvement in classification accuracy, increasing from 0.9909 to 0.9943. 
%More importantly, a substantial reduction in FRR is observed, decreasing from 0.4364 to 0.1909. This indicates that the LLM-assisted component improves the system's ability to correctly accept genuine attempts, thereby enhancing intra-user consistency handling under varying conditions. In contrast, the FAR remains low in both settings, with a minor increase from 0.0013 to 0.0023 when LLM assistance is introduced. This indicates a slight operating trade-off between FAR and FRR.
% \addnew{

\textbf{Rule-Based Mode}. Across all three Gaussian smoothing modes, rule-based mode substantially improves over the non-LLM baseline, with accuracy ranging from 99.32\%--99.46\%, FAR from 0.20\%--0.22\%, FRR from 18.27\%--26.78\%, and EER from 9.24\%--13.49\%. The Fused configuration achieves the best overall performance, with an accuracy of 99.46\%, FAR of 0.21\%, FRR of 18.27\%, and EER of 9.24\%, an improvement of 25.37 percentage points in FRR and 12.65 percentage points in EER over the non-LLM baseline, at the cost of only a 0.08 percentage point increase in FAR.

\textbf{Score-Based Mode}. Across all three Gaussian smoothing modes, score-based mode likewise substantially improves over the non-LLM baseline, with accuracy ranging from 99.22\%--99.38\%, FAR from 0.38\%--0.58\%, FRR from 13.80\%--14.40\%, and EER from 7.09\%--7.48\%. The Fused configuration again achieves the best overall performance, with an accuracy of 99.38\%, FAR of 0.38\%, FRR of 13.80\%, and EER of 7.09\%, an improvement of 29.84 percentage points in FRR and 14.80 percentage points in EER over the non-LLM baseline, at the cost of a 0.25 percentage point increase in FAR.

Comparing the two schemes' best configurations, score-based mode yields a marginally lower accuracy and a modestly higher FAR than rule-based, but achieves a substantially lower FRR and EER, reducing FRR by roughly 4.47 percentage points and EER by roughly 2.15 percentage points relative to the best rule-based configuration. This indicates that the continuous, per-user-normalized representation used in score-based is more effective at correctly accepting genuine authentication attempts, at the cost of a slightly higher rate of impostor acceptance. This suggests that score-based is preferable in usability-sensitive deployments where minimizing false rejections is a priority, while rule-based may be preferable where minimizing false acceptance is the primary concern. 
Across both schemes, the fused configuration consistently achieves the lowest EER, indicating that combining multi-scale Gaussian smoothing improves the robustness and discriminative capability of the extracted behavioral representation by jointly capturing both transient and sustained blowing patterns. While both LLM-assisted schemes yield substantial improvements
over the non-LLM baseline, they do not eliminate authentication errors entirely; false acceptances and false rejections still occur under certain conditions, as reflected in the FAR and FRR values above. Beyond these quantitative improvements, the LLM-assisted approach additionally provides a natural language justification for each authentication decision, offering a level of interpretability that is not available in the conventional threshold-based scheme. 
% }
\begin{comment}
Compared to the non-LLM scheme, %LLM-Assisted configuration, the LLM-assisted configurations 
the LLM-assisted scheme substantially improves the accuracy from 0.9909 to 0.9943 while it significantly improves the EER from 0.2189 to 0.0966. The FRR is also significantly improved from 0.4364 to 0.2647 (under the Fast configuration), 0.2580 (under the Slow configuration), and 19.09 (under the Fused configuration). This indicates that the LLM-assisted component improves the system's ability to correctly accept genuine authentication attempts, thereby enhancing intra-user consistency handling under varying conditions. Although a slight increase in FAR is observed, the FAR remains consistently low across all evaluated LLM-assisted configurations, increasing marginally from 0.0013 in the non-LLM scheme to 0.0021, 0.0023 and 0.0023 in the Fast, Slo and Fused LLM-assisted configurations, respectively. This suggests a modest trade-off between FAR and FRR while achieving significantly improved genuine-user acceptance performance. Furthermore, the Fused configuration achieves the lowest EER among the evaluated LLM-assisted configurations, indicating that combining multi-scale Gaussian smoothing improves the robustness and discriminative capability of the extracted behavioral representations by jointly capturing both transient and sustained blowing patterns.
\end{comment}

\begin{table}[htbp]
\centering
\caption{Accuracy results}
%\howard{The Without LLM is updated result. Only With LLM-assisted(rule-based) Fast is not updated} \\\eyasu{The non-LLM result is not 99.09, there is the 99.35, which we discussed about it and used in the comparison table. Or, are you considering only what we reproduced?}
%\\\howard{yes, because 99.35 is session-user result. cannot be compared directly with the accuracy result we run which is in this table are all session-session result. So the accuracy evaluation we have here are all reproduce in this LABBA paper and not including others. All this table are reproduceable in the artifacts. Not the 99.35.}
\label{tab:evaluation}
\begin{comment}
\begin{tabular}{lllll}
%\cline{1-4}
\toprule
\textbf{Method} &  \textbf{Accuracy} & \textbf{FAR} &  \textbf{FRR}  \\ \midrule
Without LLM-Assisted & 0.9909  & 0.0013 & 0.4364 &  \\ %\cline{1-4}
With LLM-Assisted    & 0.9943   & 0.0023 & 0.1909 &  \\ \bottomrule
\end{tabular}
\end{comment}
\resizebox{\columnwidth}{!}{
\begin{tabular}{clllll}
\toprule
\multicolumn{1}{l}{\textbf{Method}} & \textbf{Gaussian} & \textbf{Accuracy (\%)} & \textbf{FAR (\%)} & \textbf{FRR (\%)}  & \textbf{EER (\%)}\\ \midrule
Without LLM-Assisted & NA %\eyasu{Put "NA" under Gaussian} \howard{okay}  
& 99.09 & 0.13 & 43.64 & 21.89 \\ \midrule[0.1pt]
\multirow{3}{*}{With LLM-Assisted (Rule-Based)} & Fast  & 99.32 & 0.20 & 26.78 & 13.49 \\
                                   & Slow  & 99.33 & 0.22 & 25.20 &  12.71\\
                                   & Fused & \textbf{99.46} & \textbf{0.21} & 18.27 & 9.24\\ \midrule
\multirow{3}{*}{With LLM-Assisted (Score-Based)} & Fast  & 99.25 & 0.58 & 14.22 & 7.40 \\
                                   & Slow  & 99.22 & 0.53 &  14.4 & 7.48 \\
                                   & Fused & 99.38 & 0.38 & \textbf{13.8} & \textbf{7.09}\\ \bottomrule
\end{tabular}
}
\end{table}

\subsection{Illustrating the LLM Output with Examples}

%\howard{do we need to have 4 of these? or we just present TP and TN. Moreover, we have 2 different prompt result (old and new). Which one should we choose to discuss?}
% \addnew{
Table~\ref{tab:LLM_evaluation} presents representative examples of LLM-assisted biometric decision-making under the two evaluated modes, rule-based and score-based modes, described in Section~\ref{sec:rule_based_scoring} and Section~\ref{sec:score_based_scoring}, respectively. For each scheme, one genuine-accept and one correct-reject case are shown, allowing the LLM's reasoning to be examined along two axes: how it behaves when accepting versus rejecting an attempt, and how that reasoning differs depending on whether it is given discrete categorical labels or continuous normalized scores as input. The first two examples correspond to rule-based mode, where each channel is expressed as one of five discrete labels (VERY\_GOOD, GOOD, OKAY, WEAK, BAD), while the last two examples correspond to score-based mode, where each channel is instead expressed as a continuous similarity value on a 0--100 scale.

The table also presents the evaluation result generated by the lightweight LLM based on the corresponding input prompt. For each example, the LLM returns a single decision, ACCEPT or REJECT, accompanied by a natural language reason justifying that decision. This illustrates the interpretability the framework provides alongside its authentication outcome, as the reasoning behind each decision remains inspectable regardless of which mode, rule-based or score-based, is used to construct the prompt. 

\subsection{Attack Resilience Evaluation}
%\howard {https://www.usenix.org/system/files/usenixsecurity26-ling.pdf - USENIX 2026 paper on jailbreak attack on speech-based LLM. Can be used to support our claim in attack resilience}
The proposed LLM-assisted blow-acoustic behavioral biometric authentication framework demonstrates %strong 
inherent resilience against several common biometric attacks due to the dynamic and context-dependent nature of blow-acoustic interactions. The framework is naturally resistant to mimicry attacks because personalized blow-acoustic patterns contain complex temporal, spectral, and behavioral characteristics that are difficult to accurately imitate.

The proposed approach is also resilient against biometric template compromise attacks since it does not rely on storing conventional fixed biometric templates. Instead, authentication is performed using %dynamic behavioral interactions and contextual reasoning. 
dynamically generated and processed behavioral embeddings. In addition, the framework exhibits natural robustness against replay attacks because capturing and realistically replaying blow-acoustic interactions is highly challenging. Furthermore, modern liveness detection techniques \cite{chekole2026blowlive} can be easily integrated to further strengthen replay attack resistance.

It is also difficult %infeasible 
 to attack the framework using AI-generated synthetic acoustic sounds. Although synthetic sounds can imitate certain acoustic characteristics, they often fail to reproduce the rich contextual and behavioral information required by the LLM-based reasoning engine. As a result, the proposed framework evaluates not only signal similarity but also contextual consistency and behavioral coherence during authentication.

Compared to general-purpose LLM models, the proposed framework is relatively %inherently resilient 
less vulnerable %against 
to %model poisoning, 
prompt injection and jailbreaking attacks due to its constrained reasoning architecture. The LLM operates exclusively on structured behavioral biometric and contextual inputs generated by trusted system components, rather than unrestricted natural language prompts, %thereby significantly reducing 
which reduces the attack surface. %\addnew{
This design also mitigates a recently demonstrated class of physical-layer attacks against speech-driven LLMs, in which an adversary covertly injects jailbreak prompts as inaudible near-ultrasonic audio recovered only after microphone nonlinearity, achieving high success rates specifically against systems that process raw audio directly~\cite{ling2026sirens}. Because LABBA's LLM never receives raw acoustic input, only structured similarity representations produced by the similarity computation function $\mathcal{R}(\cdot)$, this covert acoustic injection channel does not apply to our framework. %} 
In addition, the LLM is restricted to bounded contextual authentication reasoning within controlled boundaries. Nevertheless, additional strategies, e.g., trusted training, constrained fine-tuning, and contextual consistency validation, may still be required to further enhance the robustness against adversarial manipulation.

\subsection{Usability Evaluation}

The proposed LLM-assisted blow-acoustic behavioral biometric authentication framework demonstrates strong usability in terms of ease of use, non-invasiveness, and contextual robustness. The authentication process requires only a few natural blowing interactions toward the smartphone microphone, making the system simple and convenient for users. In addition, the approach is non-invasive and contactless, as it does not require physical touch interaction with the smartphone during authentication.

The proposed framework also exhibits strong robustness under dynamic behavioral and environmental conditions. Experimental evaluations were conducted under diverse contexts, including sitting and standing postural positions, varying blowing intensities and patterns, and different phone interaction dynamics. Despite these variations, the authentication performance remained highly stable because the LLM-based reasoning engine adaptively interprets natural behavioral and contextual variability. %\howard{Not sure about this claim, we do not have the variaty of result based on these conditions to claim stability} \eyasu{This one should be fine considering the experimental dataset somewhat consists of variabilities in the mentioned contexts. In fact, we should mentioned the quit and noisy environment too as it is hardly affected by external noise.}

Furthermore, the system was evaluated under %both quiet and noisy environments. 
%mixed 
real-world acoustic conditions, involving quiet and noisy environments. The authentication performance remained reliable due to the use of band-pass filtering and Gaussian smoothing techniques, which effectively suppress environmental noise and unwanted frequency components prior to feature extraction, %\addnew{
which is intended to support reliable performance across varying acoustic environments. %} 

We acknowledge that LLM outputs are stochastic and may occasionally be unfaithful to the input (hallucination), which is a practical concern for an authentication decision. A systematic study of decision consistency across repeated runs and prompt variations is left as future work.

Overall, the proposed framework provides a practical, user-friendly, and context-aware authentication solution suitable for real-world deployment environments.

\subsection{Performance Evaluation}

Table~\ref{tab:performance_evaluation} summarizes the runtime performance of the proposed protocol, including spectral feature extraction, temporal feature extraction under different Gaussian smoothing configurations (Fast, Slow, and Fused), and LLM-assisted evaluation. In particular, the Fast configuration provides lower computational cost due to smaller Gaussian smoothing kernels, whereas the Slow configuration incurs higher runtime as a result of larger smoothing windows. The Fused configuration further increases computational cost as it jointly executes both Fast and Slow temporal processing pipelines, resulting in additional processing overhead. %\addnew{
The table also shows that LLM evaluation under score-based mode takes noticeably longer than under rule-based mode (11.604s vs. 8.237s). This is primarily attributable to prompt length: the score-based prompt includes an additional explanatory clause describing the 0--100 similarity scale, and represents each channel as a numeric ratio (e.g., \texttt{100/100}) rather than a single categorical label (e.g., \texttt{VERY\_GOOD}), resulting in a longer input sequence for the LLM to process. This reflects a further trade-off beyond the accuracy-usability trade-off discussed in Section~\ref{sec:accuracy_evaluation}: the more precise, continuous representation used in score-based mode incurs additional runtime overhead, but corresponds to the mode that also achieves the lower FRR reported in Table~\ref{tab:evaluation}, indicating that improved usability through more precise prompting comes at the cost of increased inference latency. %} 
Feature aggregation and prompt generation introduce negligible computational overhead and do not significantly contribute to the overall execution time. 

The reported execution times represent the average runtime per authentication session under the experimental setup described previously. Although LLM-assisted evaluation introduces the highest inference latency, the current implementation operates entirely on CPU without GPU acceleration or deployment-level optimization. As such, the reported runtime reflects the computational characteristics of a research prototype and may be further reduced through hardware acceleration and optimized inference strategies. %\addnew{
We acknowledge that the current end-to-end runtime, ranging from 8.237s to 11.604s depending on the scoring mode, is too slow for real-time authentication deployment regardless of which mode is used. However, the primary focus of this work is to investigate whether LLM-based contextual reasoning can meaningfully assist behavioral biometric authentication, rather than to optimize for deployment-level efficiency at this stage. Addressing runtime and usability constraints through hardware acceleration, model optimization, and inference-efficient deployment strategies is left as future work.
% }

\begin{table}[htbp]
\centering
\caption{Runtime performance results}
\label{tab:performance_evaluation}
\begin{tabular}{ll}
%\hline
\toprule
\textbf{Method}                    & \textbf{Time (s)} \\ \midrule
Spectral Feature Extraction & 0.589    \\ %\midrule
Temporal (Fast-Smoothed) Feature Extraction & 0.486   \\ %\midrule
Temporal (Slow-Smoothed) Feature Extraction & 2.350    \\ %\midrule
Temporal (Fused-Smoothed) Feature Extraction & 2.832    \\ %\midrule
Feature Aggregation         & $3.2 \times 10^{-5}$ \\
Prompt Generation           & $8.7 \times 10^{-7}$ \\
LLM Evaluation (Rule-Based)              & 8.237    \\ 
LLM Evaluation (Score-Based)     & 11.604 \\

\bottomrule
\end{tabular}
\end{table}

\begin{table*}[!htbp]
\vspace{-0.5cm}
\centering
\caption{LLM evaluation examples (\colorbox{rulebasedbg}{rule-based} and \colorbox{scorebasedbg}{score-based})}%\howard{fix this table with old and new (first 2 is old, last 2 is new. Both just shows True Positive and True Negative}}
\label{tab:LLM_evaluation}
\hspace*{-0.5cm}
%\begin{tabular}[c]{p{9.6cm}|p{5.8cm}|p{1.2cm}}
\begin{tabular}[c]{p{7.6cm}|p{6cm}|p{1.5cm}}
%\hline
\toprule
\textbf{Input Prompt} & \textbf{LLM Response} & \textbf{%Correct Evaluation
Expected} \\  \midrule
\rowcolor{rulebasedbg}
\begin{tabular}[c]{@{}p{7.6cm}@{}}
% You are a biometric verification assistant for blow audio patterns.\\
TASK: Decide ACCEPT or REJECT for a blow pattern authentication attempt.\\
- embedding similarity: VERY\_GOOD\\
- blow count similarity: VERY\_GOOD\\
- gap similarity: OKAY\\
- duration similarity: VERY\_GOOD\\
- shape similarity: OKAY\\
- amplitude similarity: VERY\_GOOD\\
- mass similarity: VERY\_GOOD\\
Return ONLY:\\
Decision: ACCEPT or REJECT\\
Confidence: 0-100\\
Reason: Describe Accept/Reject Decision
\end{tabular}
&
\begin{tabular}[c]{@{}p{6cm}@{}}
Decision: ACCEPT\\
Confidence: 95\\
Reason: The blow pattern authentication attempt has shown very good similarity in embedding, blow count, duration, amplitude, and mass. Although the gap and shape similarity are only okay, the overall high similarity in critical aspects of the blow pattern leads to a decision of ACCEPT. The gap and shape similarities are not significantly different to warrant a rejection.
\end{tabular}
&
\begin{tabular}[c]{@{}p{1.5cm}@{}}ACCEPT\end{tabular} \\ \midrule
\rowcolor{rulebasedbg}
\begin{tabular}[c]{@{}p{7.6cm}@{}}
% You are a biometric verification assistant for blow audio patterns.\\
TASK: Decide ACCEPT or REJECT for a blow pattern authentication attempt.\\
- embedding similarity: BAD\\
- blow count similarity: VERY\_GOOD\\
- gap similarity: BAD\\
- duration similarity: OKAY\\
- shape similarity: VERY\_GOOD\\
- amplitude similarity: OKAY\\
- mass similarity: OKAY\\
Return ONLY:\\
Decision: ACCEPT or REJECT\\
Confidence: 0-100\\
Reason: Describe Accept/Reject Decision
\end{tabular}
&
\begin{tabular}[c]{@{}p{6cm}@{}}
Decision: REJECT\\
Confidence: 95\\
Reason: The blow pattern authentication attempt has a significant issue with the embedding similarity, which is marked as BAD. This indicates that the audio pattern does not match the expected embedding, making it highly unlikely that the attempt is genuine. Although the blow count, shape, amplitude, and mass similarities are good, the poor embedding similarity is a critical factor that leads to the rejection of the authentication attempt.
\end{tabular}
&
\begin{tabular}[c]{@{}p{1.2cm}@{}}REJECT\end{tabular} \\ \midrule
\rowcolor{scorebasedbg}
\begin{tabular}[c]{@{}p{7.6cm}@{}}
% You are a biometric verification assistant for blow audio patterns.\\
TASK: Decide ACCEPT or REJECT for this blow pattern authentication attempt. SIMILARITY SCORES (0-100 each; 100 = matches the enrolled pattern
exactly, 0 = no resemblance; each already accounts for this user's own
normal variability):\\
- embedding similarity: 100/100\\
- blow count similarity: 100/100\\
- gap pattern similarity: 90/100\\
- duration pattern similarity: 75/100\\
- shape pattern similarity: 80/100\\
- amplitude (loudness) similarity: 95/100\\
- blow "mass" (total effort) similarity: 60/100\\
Return ONLY:\\
Decision: ACCEPT or REJECT\\
Confidence: 0-100\\
Reason: Describe Accept/Reject Decision
\end{tabular}
&
\begin{tabular}[c]{@{}p{6cm}@{}}
Decision  : ACCEPT\\
Confidence: 85\\
Reason    : The blow pattern authentication attempt has high similarity scores in embedding, blow count, gap pattern, and amplitude similarity, which indicates a strong resemblance to the enrolled pattern. Although there are some differences in duration and shape pattern similarity, the overall similarity is high, and the blow "mass" similarity is also relatively close. Therefore, the authentication attempt can be accepted with a confidence level of 85.
\end{tabular}
&
\begin{tabular}[c]{@{}p{1.5cm}@{}}ACCEPT\end{tabular} \\ \midrule

\rowcolor{scorebasedbg}
\begin{tabular}[c]{@{}p{9.6cm}@{}}
% You are a biometric verification assistant for blow audio patterns.\\
TASK: Decide ACCEPT or REJECT for this blow pattern authentication attempt. SIMILARITY SCORES (0-100 each; 100 = matches the enrolled pattern
exactly, 0 = no resemblance; each already accounts for this user's own
normal variability):\\
- embedding similarity: 0/100\\
- blow count similarity: 100/100\\
- gap pattern similarity: 10/100\\
- duration pattern similarity: 15/100\\
- shape pattern similarity: 95/100\\
- amplitude (loudness) similarity: 40/100\\
- blow "mass" (total effort) similarity: 5/100\\
Return ONLY:\\
Decision: ACCEPT or REJECT\\
Confidence: 0-100\\
Reason: Describe Accept/Reject Decision
\end{tabular}
&
\begin{tabular}[c]{@{}p{5.8cm}@{}}
Decision  : REJECT\\
Confidence: 70\\
Reason    : While the blow count and shape pattern similarities are high, indicating a strong resemblance to the enrolled pattern, the low blow "mass" similarity and gap pattern similarity suggest significant differences in the blow pattern. The overall similarity scores are not high enough to confidently accept the authentication attempt.
\end{tabular}
&
\begin{tabular}[c]{@{}p{1.2cm}@{}}REJECT\end{tabular} \\
\bottomrule

\end{tabular}
\end{table*}

\section{Related work}\label{related_work}

%\subsection{Behavioral Biometric Authentication}
%\howard{are these related works similar to blowprint and blowlive behavioral biometric?}\\
%\eyasu{This could be optional. Let's leave it for now.}\\

%\subsection{Reasoning-Based Behavioral Biometric Authentication}

%\howard{we could probably add a "Model" parameters to show what model each LLM used. I think putting the accuracy evaluation might give more problems because all are using different modalities, different dataset, and even different comparison (authentication, verification, attack detection).} 
\begin{comment}
    I have checked related works compare to "https://ieeexplore.ieee.org/stamp/stamp.jsp?arnumber=11137396" -- a summary of foundational models for biometrics, which include LLM (not all). The one we are related to is only the one which do prompt-engineering and zero-shot learning as we did not do any training or fine-tuning. I have used all that are listed in there that is suitable for us.
\end{comment}
\begin{table*}[htbp]
%\vspace{-0.5cm}
\centering
\caption{Comparison of related works}
\label{tab:comparison_related_works}
\resizebox{\linewidth}{!}{
\begin{tabular}{|l|l|cc|l|l|c|c|c|}
\hline
\multicolumn{1}{|c|}{\multirow{2}{*}{\textbf{Related Work}}} &
  \multicolumn{1}{|c|}{\multirow{2}{*}{\textbf{Modality}}} &
  \multicolumn{2}{c|}{\textbf{Accuracy}} &
  \multicolumn{1}{c|}{\multirow{2}{*}{\textbf{Usage}}} &
  \multicolumn{1}{c|}{\multirow{2}{*}{\textbf{Model}}} &
  \multirow{2}{*}{\textbf{\begin{tabular}[c]{@{}l@{}}LLM Model \\ Parameters\end{tabular}}} &
  %\multirow{2}{*}{\textbf{Context-Aware}} &
  \multirow{2}{*}{\textbf{\begin{tabular}[c]{@{}l@{}}Context- \\ Aware \end{tabular}}} &
  \multirow{2}{*}{\textbf{\begin{tabular}[c]{@{}l@{}}LLM-Reasoning \\ Based Decision\end{tabular}}} \\ \cline{3-4}
 &
   &
  \multicolumn{1}{c|}{Acc (\%)} &
  EER (\%) &
   &
   &
   &
   &
   \\ \hline
Mehta et al. \cite{mehta2026detecting} &
  Keystroke Dynamics &
  \multicolumn{1}{c|}{--} &
  -- &
  Attack Detection &
  LLaMA-3.3-Instruct &
  70B &
  $\times$ &
  $\checkmark$ \\ \hline
Almehmadi \cite{almehmadi2025cyber} &
  \begin{tabular}[c]{@{}l@{}}Keystroke Dynamics, Mouse-\\ Movement, Heart Rate\end{tabular} &
  \multicolumn{1}{c|}{--} &
  0* &
  Attack Detection &
  \begin{tabular}[c]{@{}l@{}}DeepSeek-R1-\\Distill-Qwen\end{tabular} &
  32B &
  $\checkmark$ &
  $\checkmark$ \\ \hline
Piland et al. \cite{piland2026generalist} &
  Iris Biometrics &
  \multicolumn{1}{c|}{--} &
  -- &
  \begin{tabular}[c]{@{}l@{}}Attack Detection\end{tabular} &
  \begin{tabular}[c]{@{}l@{}}Gemini 2.5 Pro\\ Llama 3.2-Vision\end{tabular} &
  \begin{tabular}[c]{@{}l@{}}N/A\\ 90B\end{tabular} &
  $\checkmark$ &
  $\checkmark$ \\ \hline
Deandres-Tame et al. \cite{deandres2024good} &
  Face Biometrics &
  \multicolumn{1}{c|}{94} &
  -- &
  Face Verification &
  GPT-4V &
  N/A &
  $\times$ &
  $\checkmark$ \\ \hline
Hassanpour et al. \cite{hassanpour2024chatgpt} &
  Face Biometrics &
  \multicolumn{1}{c|}{95.15} &
  -- &
  Face Verification &
  GPT-4V &
  N/A &
  $\times$ &
  $\checkmark$ \\ \hline
Farmanifard et al. \cite{farmanifard2024chatgpt} &
  Iris Biometrics &
  \multicolumn{1}{c|}{--} &
  -- &
  Authentication &
  GPT-4V, Gemini &
  N/A &
  $\times$ &
  $\checkmark$ \\ \hline
Rastogi et al. \cite{rastogi2025vascular} &
  Vascular Biometrics &
  \multicolumn{1}{c|}{--} &
  -- &
  Authentication &
  GPT-4V, Gemini &
  N/A &
  $\times$ &
  $\checkmark$ \\ \hline
Zerroug et al. \cite{zerroug2025automatic} &
  Iris Images &
  \multicolumn{1}{c|}{--} &
  -- &
  Biometric Enrichment &
  BERT, GPT-4, LLaMa &
  N/A &
  $\times$ &
  $\times$ \\ \hline
Nam et al. \cite{nam2026speakerllm} &
  Voice/Speaker Biometrics &
  \multicolumn{1}{c|}{97.12} &
  -- &
  Speaker Verification &
  Qwen2.5-1.5B-Instruct &
  1.5B &
  $\checkmark$ &
  $\checkmark$ \\ \hline
Halim et al. \cite{halim2025blowprint} &
  Acoustic-based Biometrics &
  \multicolumn{1}{c|}{99.35} & 9.74 &
  Authentication &
   \multicolumn{1}{c|}{--} &
  -- &
  $\times$ &
  $\times$ \\ \hline
\textbf{LABBA (ours)} &
  \textbf{Acoustic-based Biometrics} &
  \multicolumn{1}{c|}{\textbf{99.46$^\dag$/99.38$^\ddag$}} &
  \textbf{9.24$^\dag$/7.09$^\ddag$} &
  \textbf{Authentication} &
  \textbf{Phi-3 Mini Instruct} &
  \textbf{3.8B} &
  $\pmb{\checkmark}$ &
  $\pmb{\checkmark}$ \\ \hline
\end{tabular}
}
\smallskip\\
\begin{tabularx}{\textwidth}{X}
%{\footnotesize
\textbf{Description of Notations}: N/A: not applicable due to ChatGPT4 and Gemini do not publicly report their parameters; $\checkmark$: achieved; $\times$: not achieved; *: based on the ROC-AUC metric; $\dag$: rule-based; $\ddag$: score-based; --: not reported or available%}
\end{tabularx}
\end{table*}
%%\howard{comments on the attack detection of the model has not been address. we could just keep it on the related works and remove it from the comparison table 4. Moreover, the reviewers added on whether we should also covers related works on non-LLM behavioral biometric besides our blow-based. Remove this comments if needed.}

%\howard{This need to be restructured because its not LLM-based}
\subsection{Behavioral Biometrics}

Behavioral biometrics \cite{finnegan2024utility,fereidooni2023authentisense,khan2024mouse} have gained increasing attention due to their continuous and unobtrusive authentication capabilities and their potential to enhance resilience against spoofing attacks by leveraging distinctive behavioral characteristics. %\addnew{
Closer to our own modality, blowing-based acoustic patterns have recently been proposed as a behavioral biometric~\cite{halim2025blowprint}, using %Dynamic Time Warping 
DTW for fixed similarity-based verification. A follow-up work~\cite{chekole2026blowlive} extended this modality using GFCC-based spectral features with a CNN encoder, similarly relying on fixed cosine similarity for the final decision. Both works establish blow-acoustic signals as a viable behavioral biometric modality but rely on conventional, fixed similarity-based decision-making without dynamic, contextual reasoning, motivating our integration of LLM-assisted reasoning into such modality. %}

\subsection{LLM-Assisted Biometrics}

Recently, there has been increasing interest in integrating LLMs into biometric systems for tasks such as biometric detection \cite{mehta2026detecting, almehmadi2025cyber,piland2026generalist} and recognition / authentication\cite{deandres2024good,hassanpour2024chatgpt,farmanifard2024chatgpt,rastogi2025vascular,zerroug2025automatic,nam2026speakerllm} across different biometric modalities. This growing intersection between foundation models, especially LLMs, and biometrics has recently been surveyed in the literature~\cite{shahreza2025foundation}.

Mehta et al.~\cite{mehta2026detecting} explored the use of keystroke dynamics as a behavioral biometric for detecting academic dishonesty. Their work benchmarked LLaMA-3.3-70B-Instruct for identifying LLM-assisted cheating, demonstrating the potential of LLMs for behavioral biometric-based attack detection. Almehmadi~\cite{almehmadi2025cyber} introduced a framework for detecting coercion as a security state using LLM-assisted multimodal biometrics. The approach integrates behavioral signals such as keystroke dynamics and mouse movements, physiological signals including heart-rate variability, and contextual user commands to infer potential coercion scenarios. Piland et al.~\cite{piland2026generalist} extended LLM-based biometric reasoning to visual modalities by investigating iris presentation attack detection (PAD) using multimodal LLMs. Their results show that models such as Gemini and Llama-Vision can leverage pretrained visual representations and human salience-guided prompting to distinguish bona fide from attack samples without task-specific training. These studies \cite{mehta2026detecting,almehmadi2025cyber,piland2026generalist} reflect a growing trend in leveraging LLMs for reasoning and decision-making over multimodal biometric signals.

Deandres-Tame et al.~\cite{deandres2024good} investigated the effectiveness of LLMs, particularly ChatGPT, in face biometric applications, including face recognition, soft biometric attribute analysis, and explainable biometric reasoning. Similarly, Hassanpour et al.~\cite{hassanpour2024chatgpt} evaluated ChatGPT for biometric-related tasks such as face recognition, gender detection, and age estimation. Their work demonstrated that LLMs are capable of generating human-readable explanations for biometric decisions by describing distinguishing facial characteristics and auxiliary biometric traits. Similarly, Farmanifard et al.~\cite{farmanifard2024chatgpt} explored the use of LLMs, including ChatGPT and Gemini, for iris biometric authentication tasks. Furthermore, Rastogi~\cite{rastogi2025vascular} investigated the application of LLMs, particularly ChatGPT-4, in vascular biometric recognition. The study showed that ChatGPT-4 achieved strong adaptability and accuracy in vein identification tasks across multiple scenarios. These studies~\cite{farmanifard2024chatgpt,rastogi2025vascular} collectively demonstrate that LLMs can exhibit notable adaptability in biometric recognition when guided by appropriate prompt engineering strategies. However, these approaches~\cite{deandres2024good,farmanifard2024chatgpt,rastogi2025vascular} primarily rely on large-scale proprietary LLMs that require cloud-based API access, resulting in significant computational and operational costs that limit their deployment in resource-constrained environments.

%\addnew{
In the audio domain, Nam et al.~\cite{nam2026speakerllm} proposed SpeakerLLM, a speaker-specialized audio-LLM for speaker verification reasoning, achieving 97.12\% verdict accuracy on VoxCeleb1-O. Like our approach, it uses a lightweight LLM (Qwen2.5-1.5B-Instruct). However, it requires training a custom audio tokenizer so the LLM can understand speaker embeddings directly. Our approach needs no such training: biometric evidence is converted into plain text before reasoning, so an off-the-shelf LLM can be used directly through prompt engineering alone. %}

\subsection{Comparison of Related Works}
%\howard{I think we might just cut this subsection completely because discussion of the related work comparison individually has been addressed in related work section. If we want this subsection, we can just compare it directly with blowprint/blowlive result and their differences (session-session and user-based). Justify the differences in accuracy achieved, etc. The rest of related works are not giving us any benefit (Different modality, different Model, different Usage, and most of them do not present its accuracy result).} \\ \eyasu{BTW, we can also move the related work section here and merge the discussions. It is common to do so.}
%In Table~\ref{tab:comparison_related_works}, we summarize the comparison between LABBA and relevant existing works discussed above. 
To the best of our knowledge, there is limited existing work that directly investigates LLM-based behavioral biometrics. Therefore, we identify the most closely related available studies in behavioral biometric authentication and compare their approaches with our proposed work, LABBA, as summarized in Table~\ref{tab:comparison_related_works}. %We observe that 
Most existing LLM-assisted biometric studies primarily focus on physiological biometrics, particularly face, iris, and vascular image recognition and analysis. %, while behavioral biometrics have mainly been explored for attack detection or malicious activity identification rather than user authentication. 
In contrast, the use of LLMs in behavioral biometrics has largely been investigated for attack detection and malicious activity identification rather than for user authentication. %In contrast, 
LABBA explores a novel direction by introducing LLM-assisted authentication into the behavioral biometrics domain, specifically for phone-blowing acoustic behavioral biometrics. Therefore, it demonstrates the feasibility of leveraging LLM-based reasoning for behavioral biometric authentication and extends the applications of LLMs beyond the conventional physiological biometric scenarios.

In addition, although Halim et al. \cite{halim2025blowprint} is not an LLM-based work, we include it in our comparison because it employs a closely related blow-acoustic biometric modality. While their approach does not incorporate reasoning-based authentication decisions, it reports an %remarkable 
accuracy of 99.35\%, which is slightly lower than the accuracy achieved by LABBA (99.46\% with the rule-based mode and 99.38\% with the score-based mode). The reported EER of Halim et al. (9.74\%) is also higher than that of LABBA (9.24\% with the rule-based mode and 7.09\% with the score-based mode). Moreover, we note that the reported accuracy in Halim et al. \cite{halim2025blowprint} is obtained using a session-to-user comparison based on KNN, where each probe session is compared only with the nearest $n$ sessions among the available sessions of a user, rather than being evaluated through exhaustive pairwise (session-to-session) comparisons. %The former can potentially yield higher accuracy because each probe session is compared against an aggregated user representation rather than directly against individual biometric sessions. 
Such an evaluation may yield higher accuracy because %the KNN-based decision considers only the most similar sessions when determining the user's identity. 
in KNN, each probe session is compared against only the most similar sessions rather than directly against each individual biometric session. In contrast, LABBA adopts a pairwise evaluation, in which each biometric session is directly compared with other sessions. This provides a more fine-grained evaluation of the system's ability to distinguish genuine and impostor biometric sessions.

\section{Conclusion}\label{conclusion}

 In this work, we proposed an LLM-assisted behavioral biometric authentication framework that performs authentication through contextual trust evaluation rather than relying solely on a single fixed similarity threshold. The proposed framework leverages an LLM as a contextual authentication reasoner capable of integrating spectral and temporal biometric similarity descriptors to support adaptive authentication decisions. The framework is evaluated using a novel phone-blowing behavioral biometric technique that captures rich environmental and behavioral contextual information. 
 %Therefore, the behavioral biometric authentication is dynamically determined based on adaptive trust reasoning rather than rigid static threshold matching.

Experimental results demonstrate that the proposed approach improves authentication flexibility and reduces the false rejection rate while maintaining a low false acceptance rate, achieving an overall accuracy of 99.46\% and 99.38\% under rule-based and score-based modes, respectively.
%Experimental results demonstrate that the proposed approach improves authentication flexibility and reduces the false rejection rate while maintaining a low false acceptance rate, achieving an overall accuracy of 99.43\%. 
The findings establish a new direction for behavioral biometric authentication by transforming conventional signal-based verification into an intelligent cognitive authentication framework capable of adaptive, context-aware, and semantically informed security decision-making.

Despite the promising results, several directions remain for future research. First, the current implementation operates entirely on CPU-based execution, and future work may explore GPU acceleration and cloud-based deployment strategies to improve inference efficiency and scalability. Second, alternative LLM models with varying model sizes and computational requirements may be investigated to further optimize the trade-off between computational cost and authentication performance. Finally, future studies should examine security and privacy-preserving mechanisms for LLM-assisted biometric authentication, including secure feature handling, privacy-aware inference, and resistance against adversarial manipulation.
%\input{09_acknowledgments}

%\subsection*{\textbf{LLM Usage Statement}}
%LLMs were used for editorial purposes in this manuscript, and all outputs were inspected by the authors to ensure accuracy and originality.

%----------------------------------------------------------------------------------------
%	BIBLIOGRAPHY
%-------------------------------------
%References
% \bibliographystyle{ACM-Reference-Format}
\bibliographystyle{splncs04}
\bibliography{10_references}

%Appendix
\appendix

\section{Open Science}

The source code and artifacts of LABBA are available at \cite{labba_dataset}.

\end{document}